# VIOLATION OF THE INVERSE SQUARE LAW BY THE EMISSIONS OF SUPERSONICALLY OR SUPERLUMINALLY MOVING VOLUME SOURCES

H. Ardavan[1] and J.E. Ffowcs Williams[2]

[1]*Institute of Astronomy, University of Cambridge*
*Madingley Road, Cambridge CB3 0HA, U.K.*
[2]*Department of Engineering, University of Cambridge*
*Trumpington Street, Cambridge CB2 1PZ, U.K.*

The generally familiar notion that the conservation of energy requires the intensity of the radiation generated by a localized finite-duration source to decay like the inverse square of the distance from the source is not necessarily true. In this paper, we identify physically tenable sources of acoustic and electromagnetic radiations the amplitudes of whose emissions to particular distant zones decay cylindrically (like $R_P^{-\frac{1}{2}}$) rather than spherically (like $R_P^{-1}$) as $R_P$ tends to infinity ($R_P$ denotes the distance of the observer from the source). These sources have moving distribution patterns which are in general three-dimensional and which propagate faster than the emitted waves. Their emission is characterized by a waveform of constant duration that consists of a continuous assemblage of cylindrically decaying subpulses. Each subpulse embodies a propagating caustic and is narrower the further away it is observed from the source. The change in the lifetime of the subpulses with range ($\sim R_P^{-1}$) is such that their energy—but not their intensity—follows the inverse square law and the Rayleigh distance associated with them is of the same order of magnitude as their distance from the source ($R_P$) for all values of this distance.

We present our work in the context of the literature on the non-diffracting wave packets known as acoustic or electromagnetic missiles, and point out how these missiles allow the existing body of data on the emissions from supersonic jets and propellers and from pulsars to be seen in a different light. A supersonically convected aeroacoustic source of volumetric scale $L^3$ and lifetime $T$ radiates conventional Mach waves whose mean square pressure-fluctuations level scales as $\rho^2 U^4 cTL/R_P^2$ ($\rho$ and $U$ are the density and velocity of the fluid and $c$ is the speed of sound). But precisely at the Mach angle the cylindrically decaying subpulses will augment the general field via a fine-grained system of caustics in which the mean square pressure level is bigger by a factor of $R_P/L$. In the case of pulsars, where the rotating pattern of the electric charge-current distribution in the magnetosphere of the central neutron star attains superluminal phase speeds, this same system of caustics can be seen in the microstructure of the observed radio pulses. The large value of the factor by which the brightness temperature of a pulsar's radio emission exceeds the kinetic temperature of its magnetospheric plasma is compatible with that of the factor $R_P/L$, $L$ being the length scale of the emitting plasma.

## 1. Introduction

Within the past decade, a class of exact solutions of the homogeneous wave equation has been reported in the literature which are described by such graphic counter-intuitive terminology as non-diffracting radiation beams, directed energy pulse trains, or focus wave modes (Brittingham 1983, Belanger 1984, Sezginger 1985, Ziolkowski 1985, Durnin *et al.* 1987); these are essentially localized energetic centres that propagate through space without decay. Some, though possessing a finite energy density, contain an infinite total energy (Wu & Lehmann 1985) and have other properties in common with plane waves, but it is possible to construct finite-energy wave packets from them by superposition. Attempts at launching these wave packets from planar apertures of finite area seem to result in waves which, though remaining localized beyond the Rayleigh distance, eventually spread and decay spherically (Ziolkowski *et al.* 1989 & 1991; Hafizi & Sprangle 1991; Lapointe 1992).

There exists, in addition, a class of exact inhomogeneous solutions of the wave equation where the wave energy contained in a pulse is both finite and decreases more slowly than the inverse square of the distance from the localized source of the pulse—no matter how large the distance may be (Wu 1985; Ffowcs Williams & Guo 1988; Shen & Wu 1989; Myers *et al.* 1990; Shen *et al.* 1991; Ffowcs Williams 1992). These pulses, that are known as acoustic or electromagnetic missiles, *can* be generated by finite-area planar source distributions provided that the spectrum of the source has no cut-off and decreases sufficiently slowly at high frequencies (Chengli & Chenghua 1989). By a suitable choice of the behaviour of the source spectrum at high frequencies, it is apparently possible to construct solutions for which the energy carried by the pulse decays like $R_P^{-\epsilon}$ with an arbitrarily small $\epsilon$, $R_P$ being the distance of a far-field observation point $P$ from the source (Myers, Wu & Brandt 1990).

The non-spherically decaying pulses described by these solutions are obtained by invoking conditions which push the boundary between the near and far zones towards infinity, i.e. which extend the Fresnel regime of diffraction and erode the Fraunhofer zone. The Rayleigh distance $L_R = l^2/\lambda$, that marks the boundary in question, tends to infinity either when the dimension $l$ of the diffracting aperture tends to infinity or when the wavelength $\lambda$ of the radiation approaches zero. The non-diffracting beams described by the homogeneous solutions have infinitely large transverse extents $l$, and the missiles described by the inhomogeneous solutions contain vanishingly small wavelengths $\lambda$.

Of course, any source that can be realized in the laboratory has neither an infinitely wide extent nor an infinitely broad spectrum for the Rayleigh distance associated with its radiation to extend to infinity. This and other illusive requirements for simulating diffraction-free radiation beams are closely related to the difficulties encountered in the design of supergain antennas. It has been known for some time that to achieve high directivity in the beam from an antenna array of specified area, it would be necessary to introduce phase reversals within the aperture over distances that are short compared to a wavelength (Toraldo di Francia 1952; Rhodes 1974). There is a similar constraint on those planar sources whose fields simulate the non-diffracting homogeneous solutions best: both the amplitudes and the phases of the individual radiating elements within the aperture have to undergo rapid spatial variations independently of one another (see Ziolkowski 1991). The engineering problems that are posed by these requirements are so severe as to render their implementation impractical. Associated with such rapid variations in amplitude or phase is an amount of stored energy in the near field which is by orders of magnitude greater than that associated with a uniformly excited aperture. Thus, not only is an extreme precision needed to achieve the required variations over short distances within the aperture, but the ohmic losses that accompany the electric currents circulating in the antenna structure are also impractically large (Johnson & Jasick 1984).

Further progress in the field hinges on finding physically realizable sources for the reported non-diffracting solutions. The Rayleigh constraint does not, as is commonly thought, exclude the possibility of realizing such sources. This constraint stems from the uncertainty relation $\Delta x \Delta k \gtrsim 1$ that connects the spreads $\Delta x$ and $\Delta k$ in the values of any two dual variables $x$ and $k$ in a Fourier transform. In the far-field approximation where the wave amplitude and the source density are related via a Fourier transformation, the smaller the extent $\Delta x_\perp$ of an aperture, the larger is the spread $\Delta k_\perp$ in the component of the wave vector $\mathbf{k} = (k_\parallel, k_\perp)$ normal to the direction of propagation, and hence the larger is the beamwidth $\Delta\theta = \Delta k_\perp / k_\parallel \simeq \lambda \Delta k_\perp$. Thus, the distance within which the radiation beam emanating from the aperture remains collimated, i.e. the Rayleigh distance, cannot exceed $L_R \simeq$ $\Delta x_\perp/\Delta\theta \simeq (\Delta x_\perp)^2/\lambda$ (see Hafizi & Sprangle 1991). However, in situations where the amplitude of the radiated wave is no longer simply proportional to the Fourier transform of the source density, the radiation beam is not necessarily subject to spherical spreading at large distances; such situations arise when the radiation beam contains propagating caustics (cf. Ardavan 1994 a; Ffowcs Williams 1994). At a caustic, the phase function which describes the space-time distance between each source point $\mathbf{x}$ and the observation point $\mathbf{x}_P$ in a radiation integral has coalescent stationary points, and so cannot be adequately represented by the first two terms in its Taylor expansion in powers of $|\mathbf{x}|/|\mathbf{x}_P|$, as is done in the far-field approximation.

Since the constructive interference of the contributions from different parts of the source is crucial to the formation of caustics, stationary planar source distributions are not in fact the best candidates for sources of waves which remain directed at large distances. As Ziolkowski (1989) acknowledges after attempting to design the required planar sources, it would be more feasible to arrange phase relations between neighbouring portions of a source which is distributed over a volume. Moreover, the phase relations which would lead to the crowding together of the emitted wave fronts and so the focusing of rays arise naturally if the source is moving and has a speed along the radiation direction which is comparable to the wave speed.

The following properties of the homogeneous solutions in question are also revealing as far as the nature of their sources is concerned. One is that they are solutions to the Goursat problem for the wave equation, i.e. that they can be obtained by prescribing suitable Cauchy data on a moving wavefront (Hillion 1991). Such Cauchy data can in turn be mimicked by two-dimensional sources which move with the wave speed (cf. Morse & Feshbach 1953, p. 837). Another property is that these solutions transform (via the Lorentz transformation) into regular diverging wave solutions when observed in an inertial frame whose limiting speed relative to the original frame equals the wave speed (Belanger 1986). They owe their diffraction-free character, in other words, to an extreme form of relativistic beaming. A further property is that the phase speeds of the wave packets described by these solutions exceed the wave speed $c$: the amplitudes of the wave packets depend on the coordinate marking the propagation axis, $z$, and on time, $t$, in only the combination $z - ut$, where $u > c$ (see Barut 1990; Campbell & Soloway 1990; Lu & Greenleaf 1992; and Donnelly & Ziolkowski 1993). A wave field with such a symmetry ($\partial/\partial t + u\partial/\partial z = 0$) can only arise from a source which satisfies the same symmetry and so has a distribution pattern that propagates along the $z$-axis with a phase velocity exceeding $c$.

The above considerations suggest that, amongst the set of possible sources for the given solutions, there should be some that move along the line of sight of the observer with the wave speed and, as a result of this motion, give rise to a ray system which entails caustics. Certain authors (e.g. Shen *et al.* 1988; Uehara & Kirkuchi 1989; Herman & Wiggins 1991) have already noted that ray focusing must be regarded as the main criterion for whether or not a reflecting or refracting system is capable of launching missiles or generating diffraction-free beams. What does not seem to be generally acknowledged in the literature, however, is the possibility that ray focusing underlies the formation of non-spherically decaying waves by any system and can as such provide a clue for identifying the more relevant sources of these waves.

Whenever an acoustic or optical ray system has an envelope (at which adjacent wave fronts must necessarily meet at a caustic), the wave amplitude that is predicted by geometric acoustics or optics becomes infinite on this caustic. So does the field generated by certain discontinuities in the source distribution. Such a singularity is removed once the the field is constructed by means of a higher-order approximation to the exact solution, or the source is modelled more realistically. However, in cases where the envelope of the rays extends into the far field or propagates, on the caustic itself, the resulting finite value of the wave amplitude does not in general fall off as the inverse of the distance from the source. The crowding together of the wave fronts at the caustic results in a pulse whose duration is much shorter than the time interval during which it is emitted and whose strength is quite different from that which would be expected on the basis of the inverse square law.

An example of a propagating caustic which extends into the far field is encountered in the context of the Mach emission from a supersonically moving point source. The acoustic field at the Mach cone, which constitutes the caustic, is infinitely strong in this case (because the density of a point source is singular). When the source is extended, the field is everywhere finite, though still highly directed. The collection of Mach cones issuing from the various volume elements of the extended source mark out a region of space, in the form of a moving conical shell, within which the radiation field is much stronger than in other regions of space. We shall see in this paper that even when the source in question has a finite duration, the amplitude of the radiation field in the region spanned by the Mach cones decays cylindrically (i.e. like $R_P^{-\frac{1}{2}}$) rather than spherically (i.e. like $R_P^{-1}$) with the distance $R_P$ from the source. This is at the heart of our new result. We will demonstrate that the wave activity in this very special, limited zone of three-dimensional space is actually identical to the activity that would be there were the source indefinitely long-lived and radiating cylindrical Mach waves.

The source points that give rise to the cylindrically decaying field at a given observation point are those which approach the observer—along the radiation direction—with the wave speed at the retarded time. We shall show that these source points only occupy a limited part of the source whose thickness in the direction of motion is of the order of $(cT)^2/R_P$ ($T$ is the duration of the source). But, by generating waves whose fronts crowd together, they give rise to a narrow subpulse embodying a caustic which, though emitted during the entire lifetime of the source, only has a width of the order of $(cT)^2/R_P$. As the Mach cones issuing from the constituent volume elements of the source propagate past a stationary observer, the limited part of the source whose elements approach the observer with the wave speed changes. The signal detected at the next instant is another subpulse arising from the accumulation of wave fronts that are again emitted during the entire lifetime of the source, but one that does not bear any phase relationship (other than that which may be inherent in the source distribution) with the coherent signal generated by the earlier set of source points. A superposition of these narrow subpulses thus results in an overall waveform whose radial extent equals that of the source distribution along the radiation direction. Despite their cylindrical decay and their violation of the inverse square law for intensity, the individual subpulses contain an energy which still falls off with the inverse square of the distance because the slow decay ($\sim R_P^{-\frac{1}{2}}$) of their amplitudes is accompanied by the shortening ($\sim R_P^{-1}$) of their duration with range.

The fact that caustics prevent all volume elements from radiating with the same effectiveness towards a given observation point sets a supersonically moving extended source in a different category from all stationary planar source distributions. We shall demonstrate that Rayleigh's criterion for the occurrence of constructive interference between the contributions from different parts of the source distribution is satisfied in the present case at all distances from the source: the shape and size of the locus of the source elements which approach the observer along the radiation direction with the wave speed at the retarded time, i.e. which radiate most effectively towards the given observation point, is such that these elements radiate in phase automatically—irrespective of frequency, the source's extent or observer's distance. Or stated differently, the change in the width of each subpulse with range is such that its associated Rayleigh distance is of the same order of magnitude as its distance from the source for all values of this distance.

Although the flow from a supersonic rocket or jet is an example of the extended source in question, the volume sources that generate cylindrically decaying waves need not involve the bulk motion of actual fluids. Any extended source whose pattern of distribution propagates with a supersonic *phase* velocity would generate the waves described above. All that is required is that the density $s(\mathbf{x},t)$ of the source distribution should depend on a spatial coordinate, say $z$, and time $t$ in only the combination $z - ut$ while the contribution towards the subpulse is made, i.e. should have the form $s(x, y, z - ut)$. [$\mathbf{x} = (x, y, z)$ is the position vector and $u$ is a speed exceeding $c$.] The motion of a rigid body, for instance, can give rise to a flow in the medium surrounding it whose distribution is steady in the rest frame of the body. Even though the fluid motion around the body, which creates the source with the density $s(\mathbf{x}, t)$, has velocities that are small, the phase velocity of the pattern associated with the source equals the speed of the body and so can exceed the wave speed if the body moves supersonically. The distinction between a source that consists of a moving material object and a source that represents a moving pattern is not in any way reflected in the wave equation, so that the wave field which arises from a volume element of an extended moving pattern is precisely the same as that which would arise from a similar source representing a moving material element (see §2).

The subpulse fields described here have so far escaped attention in the aeroacoustic literature because it seems not to have been appreciated that supersonically convected sources *always* contribute twice to the sound at precisely the Mach angle. When a supersonic source is heard twice, waves coalesce, caustics form and the conical structure of the wave field is then inevitable [cf. Whitham (1952), Dowling & Ffowcs Williams (1983), Ffowcs Williams (1986, 1994) and Lighthill (1994)]. It may be that the separate sites of emission were believed to be too widely separated to matter and indeed that is so for the distant field away from the Mach angle. Only the very long lived sources could make multiple contributions to that. But exactly at the Mach angle associated with every supersonic source element, a caustic forms irrespective of the duration of the source, multiple contributions being made into a zone adjacent to it (Region $I_P$ of figure 5). That zone spreads only in the dimensions normal to the direction of source motion, and it is only in that zone that the inverse square law of spherical spreading is violated. At any angle finitely different from the Mach angle, cylindrically spreading pulses are absent with the field in the vicinity of the Mach angle conforming to Mach wave theory, each element of the finite-duration source having contributed only once to the instantaneous sound at a point (Ffowcs Williams 1963).

Sources that move faster than their own waves are not limited to those of sound. A familiar example is a charged particle whose speed exceeds the speed of light in a dielectric medium and so emits the Čerenkov radiation. However, just as the Mach radiation by a point source is physically unrealizable because there are no sources of sound that are strictly point-like and real fluid effects destroy sound at very high frequencies, the field of a charged particle is never discontinuous on the Čerenkov cone. The index of refraction of a real dielectric medium does not remain constant at the high frequencies at which the constructively interfering contributions towards the field at a caustic are made, so that in the case of the Čerenkov radiation the effects of dispersion are crucial at the envelope of the waves. The physically relevant electromagnetic counterparts of the acoustic sources discussed above are not sources that move in a medium; they are macroscopic electric charges and currents, in empty space, whose patterns of distribution propagate faster than light *in vacuo*.

Ginzburg (1979) has emphasized that such extended sources of the electromagnetic field are fully consistent with the requirements of special relativity: the superluminally moving pattern in their distribution is created by the coordinated motion of aggregates of subluminally moving charged particles. For a source distribution to have a density of the form $s(x, y, z - ut)$ with $u > c$, it is only necessary that the (subluminal) motion of the charged particles in the (mixed) plasma constituting the source should create a pattern—of charge separation, for instance—whose *phase* velocity ($u$) exceeds the speed of light *in vacuo* ($c$).

Superluminally moving charged patterns can arise even in familiar situations; for example, when a plane electromagnetic wave impinges obliquely on the surface of a metal. In cases where the electric field vector of the incident wave lies in the plane of incidence, a polarization charge forms at the surface of the metal whose front and alternating-sign periodic structure travel along the surface of the metal with a velocity greater than $c$ (see Bolotovskii & Ginzburg 1972). The type of source that is described by $s(x, y, z - ut)$ does not have to be specifically a moving pattern of polarization charge: any macroscopic electric current the changes in whose distribution propagate with a superluminal speed also acts as a source of cylindrically decaying electromagnetic waves.

For the sake of brevity, we shall from now on refer to the charge-current distribution patterns described above simply as 'sources'; the fact that they differ from the ordinary sources of the electromagnetic field in that they are without inertia is understood. (For a full discussion of the characteristics of such sources, see Bolotovskii & Bykov 1990 and the references therein.)

The Green's function for the electromagnetic radiation emitted by a source that moves with a constant superluminal speed along a straight line is mathematically the same as the Lienard-Wiechert potential that is encountered in the study of Čerenkov radiation—although in the former case the source moves in a vacuum and has a phase velocity that exceeds the speed of light *in vacuo*, while in the latter case the source moves in a dielectric medium and has an actual velocity that exceeds the phase velocity of light in the medium. The superluminally moving patterns that act as sources cannot be point-like (see Bolotovskii & Ginzburg 1972). So, the Lienard-Wiechert potential for the Čerenkov emission has a relevance in the present context only as the contribution of an individual volume element of an extended source. In other words, this potential acquires a significance in the case of superluminally moving sources in vacuum only after it is convolved with a volume-distributed source density.‡

The cylindrical decay of the fields generated by supersonic or superluminal sources cannot be inferred from the far-field approximation to the retarded potential that is commonly employed in radiation theory, for it arises from the terms discarded in this approximation. This slower rate of decay stems from the vanishing of the Doppler factor that appears in the denominator of the expression for the Lienard-Wiechert potential, i.e. from the higher radiation efficiency of the source points that approach the observer (along the radiation direction) with the wave speed at the retarded time. The crowding together of the wave fronts generated by such source points and the resulting formation of caustics would be overlooked if the expression for the optical distance between the observation point $(\mathbf{x}_P, t_P)$ and the source points $(\mathbf{x}, t)$ were approximated by the first two terms in its expansion in powers of $|\mathbf{x}|/|\mathbf{x}_P|$ before the integrations over the space-time trajectories of the constituent volume elements of an extended source are performed.

To extract the cylindrically decaying component of the radiation analytically, one needs to carry out a delicate calculation in which all three dimensions of the source are in general taken into account,† the radiation integrals are evaluated accurately, and the position of the observer at the time of observation coincides with that of a point within the emitted wave packet. It is nevertheless possible to understand the physical mechanisms underlying the breakdown of the inverse square law by simple geometrical arguments. We shall present a detailed account of this physical approach (in §4) after having analyzed the problem mathematically (in §§2 and 3). The geometrical arguments that lead to the cylindrical decay of the amplitude of the radiation are self-contained, so that Section 4 of the paper can be read independently of its earlier sections.

---

‡ In order not to obscure these distinctions between the conventional Čerenkov radiation and the emission considered here, we shall in this paper refer to the electromagnetic counterpart of the Mach cone simply as the 'caustic' rather than the Čerenkov cone.

† Lower dimensional source distributions can also generate fields that decay non-spherically (or even diverge), but only under special circumstances. The generic sources of non-spherically decaying waves are three-dimensional.

## 2. The radiation field of a supersonically or superluminally moving volume source of unlimited duration

We shall show in the next section that wave caustics emerge over a definite interval of time and that they are unaffected by long-term properties of their source. As a result, the fields within the subpulses that are generated by a finite-duration source are identical to the fields that would be generated in an infinite-duration case, a much more straightforward case with its cylindrical spreading being familiar from Whitham's (1952) work on the disturbances caused by supersonic projectiles. For that reason, in this section we describe the familiar cylindrical case from a three-dimensional point of view emphasizing those aspects, bestowing the subpulses' essential character, that are already present in the strictly cylindrical case.

Consider an extended source of sound (or light) waves, e.g. the air surrounding a moving object (or the continuum of electric charges and currents distributed over a volume), whose distribution pattern propagates—with no distortion—at a constant supersonic (or superluminal) speed, $u > c$, along the $z$-axis of a Cartesian coordinate system $(x, y, z)$. Under circumstances that it can be regarded as infinitely long-lived, such a source distribution has a density $s$ which depends on the coordinate $z$ and time $t$ in only the combination $z - ut$, i.e. is of the form

$$s(x, y, z, t) = s(x, y, \hat{z}), \tag{1}$$

in which

$$\hat{z} \equiv z - ut, \tag{2}$$

and $s(x, y, \hat{z})$ is an arbitrary function with finite support. The waves generated by this source are described, in the absence of boundaries, by the retarded solution

$$\psi(\mathbf{x}_P, t_P) = \int d^3x dt s(\mathbf{x}, t) \delta(t_P - t - |\mathbf{x} - \mathbf{x}_P|/c)/|\mathbf{x} - \mathbf{x}_P|, \tag{3}$$

of the wave equation

$$\nabla^2 \psi - \partial^2 \psi / \partial(ct)^2 = -4\pi s, \tag{4}$$

where $c$ is the wave speed, $\delta$ is the Dirac delta function, $(\mathbf{x}_P, t_P)$ and $(\mathbf{x}, t)$ mark the space-time positions of the observation point and the source points, respectively, and the integral in (3) extends over all space-time (see e.g. Dowling & Ffowcs Williams 1983, and Jackson 1975).

When we insert (1) in (3) and change the variables $(x, y, z, t)$ to $(x, y, \hat{z}, t)$, we find that the wave amplitude $\psi$ is given by

$$\psi(x_P, y_P, \hat{z}_P) = \int dx\,dy\,d\hat{z}\,s(x, y, \hat{z}) G_0(x, y, \hat{z}; x_P, y_P, \hat{z}_P), \qquad (5)$$

in which

$$G_0 \equiv \int_{-\infty}^{\infty} dt\,\delta(t - t_P + R/c)/R \qquad (6)$$

acts as the Green's function for this type of source, and

$$R \equiv |\mathbf{x} - \mathbf{x}_P| = \{(x - x_P)^2 + (y - y_P)^2 + [\hat{z} - \hat{z}_P + u(t - t_P)]^2\}^{\frac{1}{2}}, \qquad (7)$$

with

$$\hat{z}_P \equiv z_P - u t_P. \qquad (8)$$

Thus the radiation, too, has a moving pattern which travels along the $z_P$-axis, with no distortion, at a supersonic (or superluminal) phase speed equal to that of the source.

The argument of the delta function in (6) vanishes at the following two values of the retarded time:

$$t_\pm = t_P - c^{-1}(M^2 - 1)^{-1}\{M(\hat{z} - \hat{z}_P) \pm [(\hat{z} - \hat{z}_P)^2 - (M^2 - 1)R_\perp^2]^{\frac{1}{2}}\}, \qquad (9)$$

where

$$R_\perp \equiv [(x - x_P)^2 + (y - y_P)^2]^{\frac{1}{2}}, \qquad (10)$$

and $M \equiv u/c$. The evaluation of the integral in (6), therefore, yields

$$G_0 = \sum_{t=t_\pm} R^{-1}|1 + dR/d(ct)|^{-1}$$
$$= \frac{2\theta[\hat{z} - \hat{z}_P - (M^2 - 1)^{\frac{1}{2}} R_\perp]}{[(\hat{z} - \hat{z}_P)^2 - (M^2 - 1)R_\perp^2]^{\frac{1}{2}}}, \qquad (11)$$

in which $\theta$ is the Heaviside step function. This expression, which is also encountered in the studies of Mach waves and the Čerenkov radiation, describes the disturbance generated by a supersonically (or superluminally) moving point source. Being a retarded Green's function, $G_0$ does not possess a complete reciprocity property; it is invariant under the exchange of the source point and the observation point provided that the time-like coordinates $\hat{z}$ and $\hat{z}_P$ are in addition inverted, i.e. are respectively replaced by $-\hat{z}_P$ and $-\hat{z}$.

The factor 2 in (11) reflects the number of wavelets, emanating at differing values of the retarded time from a given source point $S$, which are received at the observation point $P$ simultaneously (see figure 1). The spherical fronts of the emitted wavelets possess an envelope on which the wavelets interfere constructively and so form a caustic. In the $(x_P, y_P, \hat{z}_P)$-space of observation points, the amplitude $G_0$ of the waves generated by a given source point $(x, y, \hat{z})$ diverges on the cone

$$\hat{z} - \hat{z}_P = (M^2 - 1)^{\frac{1}{2}} R_\perp \qquad (12)$$

that constitutes the envelope of the emitted wave fronts,† and is non-zero only inside this cone. The cone (12), which in the acoustic case is called the Mach cone, will be in this paper referred to as the caustic.

Correspondingly, for a fixed observation point $P$ in the $(x, y, \hat{z})$-space of source points the locus of singularities, and the boundary of the support, of $G_0$, i.e. the surface described by (12), consists of a cone issuing from $P$ that is the mirror image of the caustic (see figure 1). This inverted cone, which we shall henceforth refer to as the *bifurcation surface*, delineates the domain of dependence of the observation point $P$: only those source points which lie either inside or on it have caustics that enclose $P$ and so can contribute towards the value of $G_0$ at $P$.

Figure 1 here

Let us now consider an observation point $P$ for which the bifurcation surface intersects the source distribution and calculate the contribution towards the value of the wave amplitude $\psi$ from those volume elements of the source that lie in the following neighbourhood, $N$, of this surface:

$$N: \quad \hat{z}_P + (M^2 - 1)^{\frac{1}{2}} R_\perp < \hat{z} < \hat{z}_P + (M^2 - 1)^{\frac{1}{2}} R_\perp + \delta \hat{z} \qquad (13)$$

where $\delta \hat{z}$ is much smaller than the scale of variation of the source density (see figure 1). The restriction on $\delta \hat{z}$ is not essential; it is here set in order that we may approximate the source density $s$ in (5) by its value in the vicinity of the bifurcation surface, i.e. by

$$s_0(x, y) \equiv s(x, y, \hat{z})\big|_{\hat{z} = \hat{z}_P + (M^2 - 1)^{\frac{1}{2}} R_\perp}. \qquad (14)$$

---

† The amplitudes of the waves are infinitely large for a source that is point-like, but they are finite for an extended source that has a singularity-free density.

Then, according to (5), (11) and (14), the contribution of the source elements in $N$ is given by

$$\psi_N \simeq 2 \int dx dy s_0 \int_N d\hat{z} [(\hat{z} - \hat{z}_P)^2 - (M^2 - 1)R_\perp^2]^{-\frac{1}{2}}$$
$$= 2 \int dx dy s_0 \operatorname{arccosh}[1 + (M^2 - 1)^{-\frac{1}{2}} R_\perp^{-1} \delta \hat{z}]. \qquad (15)$$

When the source occupies a localized region about $x = y = 0$ and the observation point lies in the far zone, i.e. when $|x_P| \gg |x|$ and $|y_P| \gg |y|$ so that $R_\perp^2 \simeq x_P^2 + y_P^2 \equiv r_P^2$, (15) reduces to

$$\psi_N \simeq 2^{\frac{3}{2}} (M^2 - 1)^{-\frac{1}{4}} (\delta \hat{z})^{\frac{1}{2}} \Big( \int dx dy s_0 \Big) r_P^{-\frac{1}{2}}. \qquad (16)$$

That is to say, the wave amplitude $\psi_N$ decays cylindrically (like $r_P^{-\frac{1}{2}}$) with the distance $r_P$ from the trajectory of the source.†

The contribution from the source elements at the bifurcation surface decays cylindrically, rather than spherically (like $r_P^{-1}$), in part because the interval of retarded time in which such elements make their contribution is significantly longer than the interval of observation time in which their contribution is received. An observer who is stationary in the laboratory frame $(x, y, z)$ receives the signal described by $\psi_N$ during the time interval $\delta t_P = \delta \hat{z}/c$ in which the caustics issuing from the source points in $N$ propagate past him. The interval of retarded time $\delta t$ during which this signal is emitted, on the other hand, is by a factor of the order of $(r_P/\delta \hat{z})^{\frac{1}{2}}$ greater than the interval $\delta t_P$ in which it is observed. To see this, let us compare the retarded times $t$ and $t + \delta t$ at which the following two source points make their contributions: one located at $(x, y, \hat{z})$ *on* the bifurcation surface and the other located at $(x, y, \hat{z} + \delta \hat{z})$ inside this surface.

Equations (9) and (12) show that the retarded time $t$, associated with the source point on the bifurcation surface, is

$$t = t_P - c^{-1} M^{-1} (M^2 - 1)^{-\frac{1}{2}} R_\perp, \qquad (17)$$

---

† It is not difficult to see from a similar calculation that the components of the gradients of $\psi(x_P, y_P, \hat{z}_P)$, which describe the observed fields when $\psi$ is a component of the electromagnetic four-potential, also decay cylindrically (cf. Ardavan 1994 a, Appendix B).

while the retarded times $t_\pm$ associated with the point $(x, y, (M^2-1)^{\frac{1}{2}}R_\perp+\delta\hat{z})$ are

$$t_\pm = t_P - c^{-1}M^{-1}(M^2-1)^{-\frac{1}{2}}R_\perp[1 + \Delta \pm M^{-1}(2\Delta + \Delta^2)^{\frac{1}{2}}], \qquad (18)$$

where

$$\Delta \equiv (M^2-1)^{-\frac{1}{2}}R_\perp^{-1}\delta\hat{z}. \qquad (19)$$

Setting the observation point in the far zone where $R_\perp \simeq r_P$ and $\Delta \ll 1$, we find that the differences between the values (17) and (18) of the retarded time are given by

$$\delta t \simeq \mp c^{-1}(M^2-1)^{-\frac{3}{4}}(r_P\delta\hat{z})^{\frac{1}{2}}. \qquad (20)$$

Thus the ratio $|\delta t|/\delta t_P$ of the emission and the reception time intervals has a magnitude in this case that is of the order of $(r_P/\delta\hat{z})^{\frac{1}{2}}$.

The source points on the bifurcation surface, that are responsible for this effect, approach the observer—along the radiation direction—with the wave speed at the retarded time: (11) and (12) jointly yield

$$\left.\frac{dR}{dt}\right|_{\hat{z}=\hat{z}_P+(M^2-1)^{\frac{1}{2}}R_\perp} = -c. \qquad (21)$$

The motion of such source points results in a crowding together of the wave fronts emitted by them and this, in turn, gives rise to a Doppler contraction of the observed duration of the generated signal. The cylindrically decaying signal described by $\psi_N$ is detectable only by certain observers. If the observation point is located such that its bifurcation surface does not intersect, and only encloses, the source distribution, i.e. if there are no source points that approach the observer with the wave speed, then the amplitude $\psi$ does not include the contribution $\psi_N$ and so decays spherically, like $r_P^{-1}$.

The region of $(x_P, y_P, \hat{z}_P)$-space in which the amplitude of the radiation decays cylindrically is that covered by the caustics of the constituent volume elements of the source, and has the form of a conical shell whose thickness is of the order of the length scale, $L$, of the source distribution along the radiation direction. The stationary observers in the $(x_P, y_P, z_P, t_P)$-space, therefore, detect the slowly decaying signal as a propagating wave packet with the phase speed $u$ and the duration $L/c$. The waves that interfere constructively at the distance $r_P$ from the source to form this propagating caustic, are emitted, according to the preceding discussion, over a long time interval of the order of $(r_P/L)^{\frac{1}{2}}(L/c) = (Lr_P)^{\frac{1}{2}}/c$.

Mathematically, there is a cylindrically decaying component to the present radiation because the symmetry $\partial/\partial t + u\partial/\partial z = 0$ of the source density $s(x, y, \hat{z})$ transfers onto the wave amplitude $\psi$ and so reduces the dimension of the wave equation that governs the radiation by one: under this symmetry, the wave equation (4) assumes the form

$$\frac{\partial^2 \psi}{\partial x^2} + \frac{\partial^2 \psi}{\partial y^2} - \left(\frac{u^2}{c^2} - 1\right)\frac{\partial^2 \psi}{\partial \hat{z}^2} = -4\pi s(x, y, \hat{z}). \qquad (22)$$

For $u > c$, the variable $\hat{z}/u$ acts as a time-like coordinate and this reduced equation describes two-dimensional waves that propagate in the $(x, y)$-space with the speed

$$c_* = (1 - c^2/u^2)^{-\frac{1}{2}} c. \qquad (23)$$

The three-dimensional wave fronts in the $(x, y, z)$-space whose intersections with the $(x, y)$-plane travel with the speed $c_*$ are not the emitted wave fronts themselves but their envelopes; they are the caustics that issue from the various source elements (see figure 2). Each caustic propagates with the speed $c$ in the direction normal to itself, so that the rate of change of the length $l$ shown in figure 2 is $dl/dt = c$. But by virtue of forming the angle $\theta = \arcsin(c/u)$ with the $z$-axis, such a cone intersects the $(x, y)$-plane along an expanding circle whose radius $r$ increases with a supersonic (or superluminal) speed:

$$dr/dt = \sec\theta \, dl/dt = (1 - c^2/u^2)^{-\frac{1}{2}} c = c_*. \qquad (24)$$

It is only the propagating caustics of the present wave system that behave like two-dimensional waves. This is, of course, why the radiation decays cylindrically only in that region of the three-dimensional $(x_P, y_P, \hat{z}_P)$-space which is covered by the caustics of the constituent volume elements of the source.

Figure 2 here

In the above discussion, we have adopted a mathematical formalism in which the field of a moving extended source is built up from the superposition of the fields of the moving point sources that constitute it. An alternative formalism is one in which the field of the moving extended source is built up from the superposition of the fields of a fictitious set of *stationary* point sources. It is always possible to find a set of stationary point sources which are distributed in such a way as to mimic the effects of a moving source at the position of the observer at a particular observation time. To find the positions and strengths of the required point sources, we only need to rewrite the retarded solution (3) of the wave equation in its alternative form

$$\psi(\mathbf{x}_P, t_P) = \int d^3x \, s(\mathbf{x}, t_P - |\mathbf{x} - \mathbf{x}_P|/c)/|\mathbf{x} - \mathbf{x}_P|. \qquad (25)$$

For fixed values of the space-time coordinates $(\mathbf{x}_P, t_P)$ of the observer, this expression has the same structure as that which describes the field of a time-independent source with the density distribution $s(\mathbf{x}, t_P - |\mathbf{x} - \mathbf{x}_P|/c)$. Such a fictitious stationary source has a density that is different for different observers, but generates a field at the specific position $(\mathbf{x}_P, t_P)$ of the observer in space-time which is indistinguishable from that generated by the actual time-dependent source.

For certain observers, the $z$-extent of the stationary source mimicking a long-lived moving source exceeds the $\hat{z}$-extent, $L$, of the actual source by a factor of the order of $(r_P/L)^{\frac{1}{2}}$. This may be seen by estimating the extent of the stationary counterpart of the moving source given in (1). To obtain the distribution of density for the equivalent time-independent source, we must replace the variable $\hat{z}$ in $s(x, y, \hat{z})$ with

$$z - u\{t_P - c^{-1}[(z - z_P)^2 + R_\perp^2]^{\frac{1}{2}}\} = \hat{z} \qquad (26)$$

[see (2), (25) and (10)]. Solving (26) for $z$ as a function of $\hat{z}$ and $R_\perp$, we find that each moving source point $(x, y, \hat{z})$ maps onto two stationary source points $(x, y, z)$ for which

$$z = z_P + (M^2 - 1)^{-1}\{\hat{z}_P - \hat{z} \pm M[(\hat{z}_P - \hat{z})^2 - (M^2 - 1)R_\perp^2]^{\frac{1}{2}}\}. \qquad (27)$$

Let us now consider two moving source points, one on the bifurcation surface (12) and another with the same $\hat{z}$-coordinate but a distance $\Delta R_\perp (\leq L)$ inside this cone. The images of these two moving source points in the $(x, y, z)$-space of stationary source points have the $z$-coordinates

$$z_1 = z_P + (M^2 - 1)^{-1}(\hat{z}_P - \hat{z}), \qquad (28a)$$

and

$$z_{2\pm} = z_P + (M^2 - 1)^{-1}\{\hat{z}_P - \hat{z} \pm M(M^2 - 1)^{\frac{1}{2}}[R_\perp^2 - (R_\perp - \Delta R_\perp)^2]^{\frac{1}{2}}\}, \quad (28b)$$

respectively [see (27)]. For an observation point in the far zone, i.e. for $R_\perp \simeq r_P \gg \Delta R_\perp$, these yield

$$|z_{2\pm} - z_1| \simeq M(M^2 - 1)^{-\frac{1}{2}}(2r_P \Delta R_\perp)^{\frac{1}{2}}. \qquad (29)$$

The two moving source points that we have been considering, therefore, map onto points within the equivalent stationary source distribution that are a distance of the order of $(r_P \Delta R_\perp)^{\frac{1}{2}}$ apart. That is to say, the equivalent stationary source distribution has an extent in the $z$-direction that is at least $(r_P \Delta R_\perp)^{\frac{1}{2}}$ long.

The reason the resulting extent of the equivalent stationary source is by a factor of the order of $(r_P/L)^{\frac{1}{2}}$ wider than the instantaneous extent $L$ of the actual moving source is that the moving source covers the radial distance $c \times (r_P L)^{\frac{1}{2}}/c$ during the time, $(r_P L)^{\frac{1}{2}}/c$, it takes to emit the propagating caustic. From the standpoint of formulation (25) of the retarded solution of the wave equation, the field of a supersonically (or superluminally) moving source decays like $r_P^{-\frac{1}{2}}$ rather than $r_P^{-1}$ simply because of the the enormous $z$-extent of the equivalent stationary source distribution.

## 3. The radiation field of a supersonically or superluminally moving volume source of finite duration

The type of source that we have been considering cannot be regarded as infinitely long-lived (by an observer at the distance $r_P$ from the trajectory of the source) unless its duration exceeds the wave-crossing time $L/c$ by the large factor $(r_P/L)^{\frac{1}{2}}$. To take account of the fact that most physically realizable sources have life spans that are considerably shorter than $(Lr_P)^{\frac{1}{2}}/c$ when the observer is in the far zone, we shall here generalize the preceding analysis to the case in which the source density is given by

$$s(x,y,z,t) = s(x,y,\hat{z})[\theta(t) - \theta(t-T)] \tag{30}$$

and so vanishes outside the finite time interval $0 < t < T$.

Inserting the above source density in the retarded solution (3) of the wave equation and proceeding as before, we obtain

$$\psi(x_P, y_P, \hat{z}_P, t_P) = \int dx dy d\hat{z}\, s(x,y,\hat{z}) G_1(x,y,\hat{z}; x_P, y_P, \hat{z}_P, t_P), \tag{31}$$

in which the Green's function $G_1$ is given by

$$\begin{aligned} G_1 &\equiv \int_0^T dt\, \delta(t_P - t - R/c)/R \\ &= \frac{\theta[\hat{z} - \hat{z}_P - (M^2-1)^{\frac{1}{2}} R_\perp]}{[(\hat{z}-\hat{z}_P)^2 - (M^2-1)R_\perp^2]^{\frac{1}{2}}}[\theta(t_+) - \theta(t_+ - T) + \theta(t_-) - \theta(t_- - T)]. \end{aligned} \tag{32}$$

[$R$, $t_\pm$ and $R_\perp$ are the same variables as those defined in (7), (9) and (10).] Thus the finiteness of the duration of the source modifies the Green's function for the problem only by introducing the combination of the step functions involving $t_\pm$ which restrict its support.

For a given event $(x_P, y_P, \hat{z}_P, t_P)$ in the space-time of the observer, the Green's function $G_1$ is non-zero only inside a region of $(x, y, \hat{z})$-space that is bounded by the bifurcation surface (12) and the two spheres

$$t_\pm = 0: \quad (\hat{z} - \hat{z}_P - ut_P)^2 + R_\perp^2 = c^2 t_P^2, \tag{33}$$

and

$$t_\pm = T: \quad (\hat{z} - \hat{z}_P - ut_P')^2 + R_\perp^2 = c^2 t_P'^2, \tag{34}$$

where

$$t_P' \equiv t_P - T. \tag{35}$$

The specific volumes in the space of source points that are selected by the individual members of the inequalities $0 < t_\pm < T$ are as follows:

$$t_+ > 0: (\hat{z} - \hat{z}_P - ut_P)^2 + R_\perp^2 > c^2 t_P^2, \text{ and } \hat{z} - \hat{z}_P < (M - M^{-1})ct_P, \tag{36}$$

$$t_+ < T: (\hat{z} - \hat{z}_P - ut_P')^2 + R_\perp^2 < c^2 t_P'^2, \text{ or } \hat{z} - \hat{z}_P > (M - M^{-1})ct_P', \tag{37}$$

$$t_- > 0: (\hat{z} - \hat{z}_P - ut_P)^2 + R_\perp^2 < c^2 t_P^2, \text{ or } \hat{z} - \hat{z}_P < (M - M^{-1})ct_P, \tag{38}$$

$$t_- < T: (\hat{z} - \hat{z}_P - ut_P')^2 + R_\perp^2 > c^2 t_P'^2, \text{ and } \hat{z} - \hat{z}_P > (M - M^{-1})ct_P'. \tag{39}$$

The intersection of these volumes, i.e. the region allowed by the combination of the step functions in (32), is that shown in figure 3.

The black region (Region $I$) in figure 3, which lies within the cone (12) but outside the spheres (33) and (34), is the volume within the bifurcation surface in which both $0 < t_- < T$ and $0 < t_+ < T$ hold. The hatched region (Region $II$), which comprises the union of the two spheres less their intersection, denotes the volume in which only one of these two inequalities holds. The loci at which the spheres (33) and (34) either intersect one another or are tangent to the cone (12) consist of three circles normal to the plane of the figure that are here designated by $A$, $B$ and $C$. The coordinates $(\hat{z}, R_\perp)$ of these three circles are

$$( \hat{z}_A \quad \hat{z}_B \quad \hat{z}_C ) = \hat{z}_P + (M - M^{-1})c\, (\, t_P - T \quad t_P - \tfrac{1}{2}T \quad t_P \,), \tag{40}$$

and

$$\begin{pmatrix} R_{\perp A} \\ R_{\perp B} \\ R_{\perp C} \end{pmatrix} = (1 - M^{-2})^{\frac{1}{2}} c \begin{pmatrix} t_P' \\ [t_P t_P' - \tfrac{1}{4}(M^2 - 1)T^2]^{\frac{1}{2}} \\ t_P \end{pmatrix}. \tag{41}$$

Regions $I$ and $II$ of the $(x, y, \hat{z})$-space constitute the largest set of source points whose contributions—from the retarded times $0 < t < T$—can reach the observation point $(x_P, y_P, \hat{z}_P)$ at the observation time $t_P$. Within this domain of dependence of the observation point, we can write (32) as

$$G_1 = [(\hat{z} - \hat{z}_P)^2 - (M^2 - 1)R_\perp^2]^{-\frac{1}{2}} \begin{cases} 2 & \text{in Region } I, \\ 1 & \text{in Region } II. \end{cases} \quad (42)$$

The factor 2 by which the value of $G_1$ changes from one region to the other reflects the number of waves that are simultaneously received at $P$ from the source points in these regions. The source points which lie within Region $I$ influence the field at $P$ via two waves, the waves emitted at $t_\pm$. The source points within Region $II$, on the other hand, influence the field at $P$ at a single instant of retarded time ($t_+$ or $t_-$). Comparing (42) with the Green's function $G_0$ studied earlier [cf. (11)], we can see that $G_1$ and $G_0$ are identical within Region $I$. That is to say, each individual source element within Region $I$ generates a field that is indistinguishable from the field which is generated by a corresponding volume element of an infinite-duration source.

Figure 3 here

As depicted in figure 3, it is always possible to choose the observation point $P$ in such a way that Region $I$ of its domain of dependence intersects the source distribution. If we mark the $(x, y, \hat{z})$-space of source points by a coordinate system whose origin lies within the source distribution, then one such observation point is that whose bifurcation surface passes through $x = y = \hat{z} = 0$, i.e. for which

$$\hat{z}_P = -(M^2 - 1)^{\frac{1}{2}} r_P \quad (43)$$

[see (12)]. The spherical wave front emitted by the source point $x = y = \hat{z} = 0$ at the retarded time $t = 0$ arrives at such an observation point at the instant

$$t_P = c^{-1}(1 - M^{-2})^{-\frac{1}{2}} r_P, \quad (44)$$

for (43) and (44) jointly yield $R_P^2 = r_P^2 + z_P^2 = c^2 t_P^2$. An observer whose space-time coordinates satisfy (43) and (44) is in a position to receive single-particle contributions that are identical to those of infinitely long-lived point sources.

However, the fact that the Green's functions $G_0$ and $G_1$ are identical within Region $I$ does not by itself imply that the overall field of the collection of the source elements that are located in this region decays cylindrically; the volume of the part of the source distribution which lies within Region $I$ shrinks to zero as $R_P$ tends to infinity. According to (12), (33) and (34), the boundaries of Region $I$ are given by

$$\hat{z} = \hat{z}_< \equiv \hat{z}_P + (M^2 - 1)^{\frac{1}{2}} R_\perp \qquad (45)$$

and

$$\hat{z} = \hat{z}_> \equiv \hat{z}_P + \begin{cases} ut_P - (c^2 t_P^2 - R_\perp^2)^{\frac{1}{2}} & \text{if } R_{\perp A} < R_\perp < R_{\perp B}, \\ ut_P' - (c^2 t_P'^2 - R_\perp^2)^{\frac{1}{2}} & \text{if } R_{\perp B} < R_\perp < R_{\perp C}, \end{cases} \qquad (46)$$

so that the $\hat{z}$-thickness, $\hat{z}_> - \hat{z}_<$, of this region at a given $R_\perp$ (see figure 3) has the maximum value

$$\text{Max}(\hat{z}_> - \hat{z}_<) = [ut_P - (c^2 t_P^2 - R_\perp^2)^{\frac{1}{2}} - (M^2 - 1)^{\frac{1}{2}} R_\perp]_{R_\perp = R_{\perp B}}$$
$$= (M - M^{-1}) c \{ t_P - \tfrac{1}{2} T - [t_P t_P' - \tfrac{1}{4}(M^2 - 1) T^2]^{\frac{1}{2}} \} \qquad (47)$$

[cf. (41)]. The length $\text{Max}(\hat{z}_> - \hat{z}_<)$ vanishes—like $cT^2/t_P$—as $t_P$, and hence $R_P$, tend to infinity. The projection of Region $I$ onto the $(x,y)$-plane, moreover, consists of an annulus with the inner and outer radii $R_{\perp A}$ and $R_{\perp C}$ and the thickness

$$R_{\perp C} - R_{\perp A} = (1 - M^{-2})^{\frac{1}{2}} cT \qquad (48)$$

(see figure 3). So, a source distribution of length scale $L$ intersects Region $I$ of the domain of dependence of an observation point satisfying (43) and (44) across a volume that is of the order of $Lc^2T^3/t_P$ when $L > cT$ and of the order of $L^2 cT^2/t_P$ when $L < cT$.

To ascertain that the finite-duration source (30) also generates a cylindrically decaying signal, we need to show that the overall field

$$\psi_I = 2 \int dx dy \int_{\hat{z}_<}^{\hat{z}_>} d\hat{z} \, s(x, y, \hat{z}) [(\hat{z} - \hat{z}_P)^2 - (M^2 - 1) R_\perp^2]^{-\frac{1}{2}} \qquad (49)$$

that arises from the collection of the source elements within Region $I$ [see(31) and (42)] decays like $R_P^{-\frac{1}{2}}$ as $R_P \to \infty$ despite the fact that $\hat{z}_> - \hat{z}_<$ is by a factor of the order of $T/t_P$ smaller than the interval $\delta \hat{z}$ which appears in the corresponding expression for an infinite-duration source [cf. (15)]. [The integrations with respect to $x$ and $y$ in (49) extend over the projection of Region $I$ onto the $(x, y)$-plane.]

To show this, let us begin by replacing the source density $s$ in (49) with its Fourier representation

$$s(x, y, \hat{z}) = (2\pi)^{-1} \int_{-\infty}^{+\infty} dk \exp(ik\hat{z}) \tilde{s}(x, y, k). \tag{50}$$

Equation (49) then assumes the form

$$\tilde{\psi}_I = 2 \int dx dy \tilde{s}(x, y, k) \int_{\hat{z}_<}^{\hat{z}_>} d\hat{z} [(\hat{z} - \hat{z}_P)^2 - (M^2 - 1)R_\perp^2]^{-\frac{1}{2}} \exp[ik(\hat{z} - \hat{z}_P)], \tag{51}$$

where $\tilde{\psi}_I$ is the Fourier transform of $\psi_I$:

$$\tilde{\psi}_I \equiv \int_{-\infty}^{+\infty} d\hat{z}_P \exp(-ik\hat{z}_P) \psi_I. \tag{52}$$

By changing the variables of integration from $(x, y, \hat{z})$ to $x$, $y$ and

$$\zeta \equiv \operatorname{arccosh} \frac{\hat{z} - \hat{z}_P}{(M^2 - 1)^{\frac{1}{2}} R_\perp}, \tag{53}$$

we can now rewrite (51) as

$$\tilde{\psi}_I = 2 \int dx dy \tilde{s} \int_0^\epsilon d\zeta \exp(i\chi \cosh \zeta) \tag{54}$$

in which

$$\chi \equiv (M^2 - 1)^{\frac{1}{2}} k R_\perp, \tag{55}$$

$$\epsilon \equiv \operatorname{arccosh}\left[1 + \frac{\hat{z}_> - \hat{z}_<}{(M^2 - 1)^{\frac{1}{2}} R_\perp}\right] \tag{56}$$

[see (45)], and the $(x, y)$-integration extends over the projection of Region $I$ onto the $(x, y)$-plane as before. Note that the parameters $\chi$ and $\epsilon$ of the $\zeta$-quadrature in (54) are of the order of $kR_P$ and $T/t_P$ (or $cT/R_P$), respectively [cf. (10) and (47)].

A suitable method for calculating the asymptotic value of the $\zeta$-quadrature in (54) for large $R_P$ and arbitrary $k$ is the method of steepest descents (see, e.g. Bender & Orszag 1978), since this method yields the first two leading terms in the asymptotic expansion of the above integral—which represent the contributions from the stationary point $\zeta = 0$ of the phase $\chi \cosh \zeta$ of the integrand and from the boundary point $\zeta = \epsilon$ of the domain of integration—simultaneously.

To apply the method of steepest descents, we regard $\zeta$ as a complex variable and substitute $\zeta = \xi + i\eta$ in the integrand of (54) to find the real and imaginary parts of the argument of the exponential function in this integrand:

$$i\chi \cosh \zeta = \chi(-\sinh \xi \sin \eta + i \cosh \xi \cos \eta). \tag{57}$$

The constant-phase contours of steepest descents that pass through the saddle point $\zeta = 0$ and the boundary point $\zeta = \epsilon$ are therefore the contours $C_1$ and $C_2$ of figure 4 which have the equations

$$\cosh \xi \cos \eta = 1, \qquad 0 \leq \xi \leq \infty, \quad 0 \leq \eta \leq \tfrac{1}{2}\pi, \tag{58}$$

and

$$\cosh \xi \cos \eta = \cosh \epsilon, \qquad 0 \leq \xi \leq \infty, \quad 0 \leq \eta \leq \tfrac{1}{2}\pi, \tag{59}$$

respectively. Now, denoting the contour that runs along the real axis from $\zeta = 0$ to $\zeta = \epsilon$ by $C$ (see figure 4) and using the fact that $\exp(i\chi \cosh \zeta)$ is analytic within the closed domain delineated by $C_1, C_2$ and $C$, we write the original integral as

$$\int_C d\zeta \exp(i\chi \cosh \zeta) = \left(\int_{C_1} - \int_{C_3}\right) d\zeta \exp(i\chi \cosh \zeta). \tag{60}$$

The asymptotic values, for large $\chi$, of the two integrals on the right-hand side of (60) can next be found by means of Laplace's method.

Figure 4 here

Along the contour $C_1$, we have

$$\zeta = \operatorname{arccosh}(\sec \eta) + i\eta \tag{61}$$

[see (58)], and hence

$$\int_{C_1} d\zeta \exp(i\chi \cosh \zeta) = \exp(i\chi) \int_0^{\pi/2} d\eta (\sec \eta + i) \exp(-\chi \sin^2 \eta \sec \eta) \tag{62}$$

[see (57)]. It is only the immediate neighbourhood of $\eta = 0$, in which the argument of the exponential function in the integrand is maximum, that contributes to the full asymptotic expansion of this integral (see Bender & Orszag 1978). That is, we may approximate the functions appearing in the

integrand of (62) by the leading terms in their Taylor expansions about $\eta = 0$ and replace the upper limit of integration with $\infty$ to obtain

$$\int_{C_1} d\zeta \exp(i\chi \cosh \zeta) \sim 2^{\frac{1}{2}} \exp[i(\chi + \tfrac{1}{4}\pi)] \int_0^\infty d\eta \exp(-\chi \eta^2)$$
$$\sim (\tfrac{1}{2}\pi/\chi)^{\frac{1}{2}} \exp[i(\chi + \tfrac{1}{4}\pi)], \quad \chi \to \infty. \tag{63}$$

Thus the integral over $C_1$, which comprises the contribution from the saddle point $\zeta = 0$, decays like $R_P^{\frac{1}{2}}$ as $R_P \to \infty$ irrespective of the value of the wave number $k$ [see (55)].

Along the contour $C_3$, we have

$$\zeta = \operatorname{arccosh}(\cosh \epsilon \sec \eta) + i\eta \tag{64}$$

[see (59)], and hence

$$\int_{C_3} d\zeta \exp(i\chi \cosh \zeta) = \exp(i\chi \cosh \epsilon) \int_0^{\pi/2} d\eta [\cosh \epsilon \tan \eta (\cosh^2 \epsilon - \cos^2 \eta)^{-\frac{1}{2}}$$
$$+ i] \exp[-\chi \sin \eta (\cosh^2 \epsilon \sec^2 \eta - 1)^{\frac{1}{2}}] \tag{65}$$

[see (57)]. Once again, since the maximum of the exponent that appears in the above expression occurs at $\eta = 0$ for all $\epsilon$, it is only the neighbourhood of $\eta = 0$ that contributes to the asymptotic expansion of this integral for large $\chi$ and arbitrary $\epsilon$. Expanding the coefficient and argument of the exponential function in (65) in powers of $\eta$ and extending the range of integration in this expression to infinity, we can therefore write

$$\int_{C_3} d\zeta \exp(i\chi \cosh \zeta) \sim i \exp(i\chi \cosh \epsilon) \int_0^\infty d\eta \exp\{-\chi \sinh \epsilon$$
$$\times [\eta + \tfrac{1}{2}(\coth^2 \epsilon - \tfrac{1}{3})\eta^3]\}, \quad \chi \to \infty. \tag{66}$$

Had $\epsilon$ remained finite in the limit $\chi \to \infty$, it would have been sufficient to keep only the first term in the argument of the exponential function in (66); the asymptotic value of the integral, as can easily be verified, would have then decayed like $\chi^{-1}$. However, in the present case where $\epsilon$ is of the order of $\chi^{-1}$ and so vanishes in the limit [see (55) and (56)], it is essential that we retain the second term in this expression: the first term on its own would predict a non-decaying value for the integral, a prediction which would be in contradiction both with the Riemann-Lebesgue lemma and with the fact that this integral decays like $\chi^{-\frac{1}{2}}$ when $\epsilon$ has the smaller value zero [see (63)] and like $\chi^{-1}$ when $\epsilon$ has a larger value of the order of unity.

Taking account of the fact that $\epsilon = O(\chi^{-1})$ by replacing the hyperbolic functions in (66) with their values for $\epsilon \ll 1$ [see (56) and (47)], and introducing the variable $\tau \equiv (\frac{3}{2}\chi/\epsilon)^{\frac{1}{3}}\eta$ to put the resulting integral in its canonical form, we arrive at

$$\int_{C_3} d\zeta \exp(i\chi \cosh\zeta) \sim i(\tfrac{2}{3}\epsilon/\chi)^{\frac{1}{3}} \exp(i\chi) \int_0^\infty d\tau \exp[-\tfrac{1}{3}\tau^3 - (\tfrac{2}{3}\epsilon^4\chi^2)^{\frac{1}{3}}\tau]$$
$$\sim \tfrac{1}{3}i\Gamma(\tfrac{1}{3})(2\epsilon/\chi)^{\frac{1}{3}} \exp(i\chi), \quad \epsilon = O(\chi^{-1}), \chi \to \infty, \tag{67}$$

[see formulae (10. 4. 44) and (10. 4. 4) of Abramowitz & Stegun (1970)]. The integral over $C_3$, which constitutes the contribution from the boundary point $\zeta = \epsilon$, therefore decays like $R_P^{-\frac{2}{3}}$ as $R_P \to \infty$ for all $k$ [see (55)].

Equations (67), (63) and (60) jointly imply that the leading term in the asymptotic expansion of the $\zeta$-quadrature in (54) is given by

$$\int_0^\epsilon d\zeta \exp(i\chi \cosh\zeta) \sim (\tfrac{1}{2}\pi/\chi)^{\frac{1}{2}} \exp[i(\chi + \tfrac{1}{4}\pi)], \qquad \epsilon = O(\chi^{-1}), \quad \chi \to \infty, \tag{68}$$

for the contribution from the boundary point $\zeta = \epsilon$ of this integral is by a factor of the order of $(\epsilon^2\chi)^{\frac{1}{6}} = O(R_P^{-\frac{1}{6}})$ smaller than that from the stationary point of its phase, $\zeta = 0$. The substitution of (68) in (54) now yields

$$\tilde{\psi}_I \sim (2\pi)^{\frac{1}{2}}(M^2-1)^{-\frac{1}{4}}k^{-\frac{1}{2}}\exp(\tfrac{1}{4}i\pi)\int dx dy \tilde{s} R_\perp^{-\frac{1}{2}} \exp[i(M^2-1)^{\frac{1}{2}}kR_\perp]$$
$$\sim (2\pi)^{\frac{1}{2}}(M^2-1)^{-\frac{1}{4}}(kr_P)^{-\frac{1}{2}}\exp\{i[\tfrac{1}{4}\pi + (M^2-1)^{\frac{1}{2}}kr_P]\}$$
$$\times \int dx dy \tilde{s}(x,y,k), \qquad r_P \to \infty, \tag{69}$$

where use has been made of the fact that, since the source is localized about the origin, we have $|x| \ll |x_P|, |y| \ll |y_P|$, and hence $R_\perp \simeq (x_P^2 + y_P^2)^{\frac{1}{2}} = r_P$ throughout the support of $\tilde{s}(x,y,k)$ [see (10)]. The remaining integral in (69), which has to be performed over the projection of Region $I$ onto the $(x,y)$-plane, is independent of $r_P$; it is of the order of $cTL\tilde{s}(0,0,k)$ if $cT < L$ and of the order of $L^2\tilde{s}(0,0,k)$ if $cT > L$, where $L$ is the length scale of the source distribution [see (48)].

Thus the part of the source within Region $I$ of the domain of dependence of an observation point satisfying (43) and (44) (see figure 3) makes a contribution towards the Fourier transform of the field in the far zone which has the order of magnitude $|\tilde{\psi}_I| \sim |\tilde{s}|cTL(kr_P)^{-\frac{1}{2}}$, i.e. which decays cylindrically, irrespective of the size of the wave number $k$. The contribution from

the part of the source which lies within the domain of dependence of the observation point but away from the bifurcation surface has a magnitude of the order of $|\tilde{s}|cTL(kr_P)^{-1}$ [see (75) below and note that $k \sim L^{-1}$], and so is by the factor $(kr_P)^{-\frac{1}{2}}$ smaller than the cylindrically decaying contribution described by (69).‡ In contrast to this spherically decaying contribution which spreads over a wide solid angle, however, the signal described by (69) is beamed into a narrow $\hat{z}_P$-interval of the order of $cT^2/t_P$, because the $\hat{z}_P$-width of the cylindrically decaying signal is of the same order of magnitude as the $\hat{z}$-thickness of Region $I$ given in (47) (see below).†

The wavelets that are emitted by any given source point $S$ during the time interval $0 < t < T$ occupy a region of the $(x_P, y_P, \hat{z}_P)$-space that is bounded by a section of the caustic issuing from the source point and two spherical wave fronts tangent to this cone (see figure 5). The part of this region bordering on the caustic which is covered by two sets of wavelets has the same shape and dimensions as those of Region $I$ in figure 3 and will be here referred to as Region $I_P$. The waves that are received in this limited region of space are identical to those which would have been received had the source been an infinitely long-lived one; they comprise the particular set of waves contributed by each source element whose superposition leads to a cylindrically decaying signal, for the condition $\zeta = 0$ entering the derivation of (69) is only met on the caustic that is generated by a source point [see (53) and (12)]. The locus of source points the Regions $I_P$ of whose wave systems intersect the observation point $P$ is Region $I$ of figure 3. It follows, therefore, that the region of $(x_P, y_P, \hat{z}_P)$-space in which the cylindrically decaying contribution $\psi_I$ is detectable has a thickness in the $\hat{z}_P$-direction that is of the same order of magnitude as the $\hat{z}$-thickness of Region $I$, i.e. that an observer in the far zone detects the contribution $\psi_I$ as a subpulse with the exceedingly short duration $\Delta t_P \sim cT^2/r_P$ [see (44) and (47)]. The narrowness of the width of this subpulse is in fact intimately related to the cylindrical decay of its amplitude: unless the amplitude of a subpulse of duration $cT^2/r_P$ decays like $r_P^{-\frac{1}{2}}$, the wave energy contained in it would not have the dependence $r_P^{-2}$.

---

‡ In the case where $\psi$ represents the sound pressure from a supersonically convected aeroacoustic source with the density $\rho$ and velocity $U$, the effective value of the source density appearing in (69) scales as $s \sim \rho U^2 L^{-1}(cTL)^{-\frac{1}{2}}$ (see Ffowcs Williams 1963).

† The reason the angular distribution of the signal is not exhibited by (69) is that this expression was derived for a specific type of observation point; (69) only applies to an observation point the Region $I$ of whose domain of dependence lies wholly within the source distribution.

Figure 5 here

The reason the wave amplitude (69) for a source of finite duration decays in the same way as does the wave amplitude (16) for an infinitely long-lived source is that the finiteness of the duration of the source is reflected entirely in the width of the signal that is generated. There are two distinct coherence mechanisms responsible for the cylindrical decay of the signal from a volume source. One is the Doppler contraction of the reception time interval: the duration of the observed signal is by a factor of the order of $(cT/r_P)^{\frac{1}{2}}$ shorter than its emission time interval $T$. The other is the constructive interference of the contributions of the source elements which lie on the bifurcation surface: those source elements that are represented by the saddle point $\zeta = 0$ in the above analysis make a contribution towards the value of the integral in (68) which is significantly stronger than that of the source elements represented by $\zeta = \epsilon$, even when $\zeta = \epsilon$ lies at a vanishingly small distance from $\zeta = 0$. The first mechanism on its own cannot give rise to a cylindrically decaying signal unless the duration of the source is unlimited. However, since the area across which the bifurcation surface intersects a volume-distributed source is finite even when the source is short-lived (it is of the order of $cTL$ instead of $L^2$), the second mechanism is equally effective in the case of a short-lived as in that of an infinitely long-lived source. The finiteness of the duration of the source makes a radical change to the dimensions of the contributing part of the source only in the $\hat{z}$-direction (see figure 3). This change, which is effected by the first mechanism [cf. (20) for $\delta t = T$], influences the duration of the resulting signal but not its amplitude. The field in certain zones would decay cylindrically, whether the source is of finite or of infinite duration, provided that a dense set of source points lies on the locus of singularity of the Green's function $G_1$, i.e. on the bifurcation surface $\zeta = 0$.

Note that had we applied the method of stationary phase—instead of the method of steepest descents—to the integral in (68), we would have had to require the wavelength of the radiation to be much shorter than the Rayleigh distance $(cT)^2/R_\perp$ in order to arrive at the asymptotic relation given in this equation [see Appendix A of Ardavan (1994a)]. This, however, is a spurious requirement that stems from the inapplicability of the method of stationary phase to an integral whose range of integration vanishes in the asymptotic limit. The parameter $\chi$ of the asymptotic expansion considered in this section is large either when $R_\perp$ is large and $k$ is arbitrary or when $k$ is large and $R_\perp$ is arbitrary [see (55)]. When $k \gg R_\perp/(cT)^2$, it is sufficient to keep only the first term in the expansion of the argument of the exponential function in (66) because $\epsilon\chi$ remains large in the limit $R_\perp \to \infty$. Consequently, the contribution from the boundary $\zeta = \epsilon$ of the integration domain behaves like $\chi^{-1}$, just as it does in cases where the range of integration is independent of the expansion parameter. On the other hand, if $k$ is arbitrary and $R_\perp$ is large, as assumed here, we know from (67) that the second-order term in the asymptotic expansion of the above integral behaves like $\chi^{-\frac{2}{3}}$ rather than like $\chi^{-1}$. It is the inability of the method of stationary phase to handle a correction term of this type—with a fractional order—that leads to the spurious restriction on the wave number $k$.

Note also that had we started with the alternative form (25) of the retarded solution to the wave equation rather than with (3), and regarded the integrand of (25) as a classical and not a generalized function, we would have been led to expect that the far-field value of $\psi$ for a source of finite duration with volume $V$ should be of the order of $sV/R_P$. We can show, on the other hand, that it does not make any difference to the outcome of the analysis presented in this section whether we start with (3) or with (25). The discrepancy between the above order-of-magnitude estimate and the outcome of the exact analysis evidently stems from regarding the far-field limit of the integrand in (25) as a non-singular classical function. To see why it makes a difference whether we regard the integrand of the retarded potential as a generalized or as a classical function, let us consider (25) for the case in which the source has a density of the form (30) and a $z$-extent of length $L$. The $z$-quadrature in (25) can in this case be written as

$$\int_0^L dz \frac{s[x, y, z - u(t_P - |\mathbf{x} - \mathbf{x}_P|/c)]}{|\mathbf{x} - \mathbf{x}_P|}$$
$$\times [\theta(t_P - |\mathbf{x} - \mathbf{x}_P|/c) - \theta(t_P - T - |\mathbf{x} - \mathbf{x}_P|/c)]$$
$$= \int_0^{L/R_P} dz' \frac{R_P s[x, y, R_P z' - u(t_P - |\mathbf{x} - \mathbf{x}_P|/c)]}{|\mathbf{x} - \mathbf{x}_P|}$$
$$\times [\theta(t_P - |\mathbf{x} - \mathbf{x}_P|/c) - \theta(t_P - T - |\mathbf{x} - \mathbf{x}_P|/c)], \quad (70)$$

where $z' \equiv z/R_P$. If we regard the integrand on the right-hand side of (70) as a classical function, then we would expect this integral to vanish in the limit $R_P \to \infty$ at least as rapidly as its range $L/R_P$ does. But if we regard the limiting value of the same integrand as a generalized function with a vanishing support, then the extent of the range of integration becomes irrelevant. The mere fact that in the problem we have been analyzing in this section the value of the integral itself decreases more slowly with $R_P$ ($\sim R_P^{-\frac{1}{2}}$) than does the volume of the domain of integration ($R_P^{-1}$) implies that the integrand of (25) for $R_P \to \infty$ is a function with vanishing support which here acts as a distribution [see (47) and (69)].

Note, finally, that the part of the source which generates the subpulse received at $P$ does not—like ordinary sources—have an identity that is independent of the observation point $P$. Any change in the position of the observation point $P$ changes the part of the source, falling within Region $I$, that gives rise to the contribution $\psi_I$. The set of observation points Regions $I$ of whose domains of dependence intersect the source distribution span an interval in $\hat{z}_P$ equal to the $\hat{z}$-extent, $L$, of the source distribution (see figure 3). Correspondingly, the set of observation points at which subpulses are received span a pulse window whose width equals the time $L/c$ taken by a wave to cross the source. The superposition of the subpulses within this pulse window thus results in an overall waveform which, as in the case of an infinitely long-lived source, has the duration $L/c$ (see figure 10).

In contrast to the signal generated by an infinitely long-lived source, however, the present signal can have a highly uneven and rugged waveform. We have already seen (in §2) that the wavelets that interfere constructively to form the caustics issuing from the constituent volume elements of a source distribution are emitted during an interval of retarded time that is by the factor $(r_P/\delta\hat{z})^{\frac{1}{2}}$ greater than the time interval $\delta\hat{z}/c$ in which a stationary observer receives them. Though arising from a limited part of the source with a $\hat{z}$-extent of the order of $\hat{z}_> - \hat{z}_< \simeq (cT)^2/r_P$ [see (47)], the contributions that go into forming a subpulse are thus made during the entire lifetime $[r_P^{\frac{1}{2}}(\hat{z}_> - \hat{z}_<)^{-\frac{1}{2}} \times (\hat{z}_> - \hat{z}_<)/c \simeq T]$ of the source. By virtue of approaching the observer along the radiation direction with the wave speed at the retarded time, the source points within Region $I$ of the domain of dependence of the observation point generate a narrow signal embodying a caustic, which though emitted during the finite interval of retarded time, $T$, only has a duration of the order of $(\hat{z}_> - \hat{z}_<)/c \simeq cT^2/r_P$. As the source moves past a stationary observer, and so the position of the observation point in the $(x_P, y_P, \hat{z}_P)$-space changes, the volume elements of the source inside Region $I$ of the domain of dependence of the observation point also change. The narrow signal detected at the new observation point is another caustic arising from the accumulation of the wave fronts that are emitted during the entire lifetime of the source, but does not bear any phase relationship (other than that which may be inherent in the source distribution) with the coherent signal that was generated by the earlier set of source points. The overall waveform received by the stationary observer thus consists of the incoherent superposition of the signals of numerous coherent emitters (see figure 10).

Since the site of emission of each subpulse is determined by the position of the observation point itself, the subpulse structure of the amplitude of the overall waveform has a pattern that changes radically as the distance $r_P$ changes. This change, in conjunction with the fact that the subpulses are narrower the further away they are observed from the source, renders the
constant width of the overall waveform compatible with the conservation of
energy. In general, no two subpulses within the waveforms that are observed
at different distances have identical sources or are composed of the same
set of propagating waves, so that the $r_P$-dependence of the amplitude or
width of an individual subpulse is not well-defined. Nor is it possible to
identify a unique source distribution for the different overall waveforms that
are observed at different distances. However, the reduction in the widths of
the subpulses and the change in the subpulse structure of the waveform with
increasing distance from the source alter the distribution of the flux density
of energy over the cross-sectional area of the overall radiation beam in such
a way that the integral of this flux density over the area in question becomes
independent of $r_P$. That this should be so is dictated by the fact that the
retarded solution of the wave equation—from which the cylindrical decay of
the amplitude of the subpulses is derived—is manifestly compatible with the
conservation of energy as far as the wave energy emitted by the entire source
is concerned.

## 4. Physical interpretation of the results

Now that the physical processes responsible for the coherence of the
generated signal are pinpointed, essentially all of the above results can be
obtained—to within an order of magnitude—by means of simple geometrical
arguments. In this section we present these arguments in some detail both
to elucidate the physical content of the above analysis itself and to expound
the basic mechanisms that underlie the unexpected features of its outcome.

Consider a localized source of sound such as that depicted in figure 6.
Imagine that it is steady in its own rest frame and that it has been moving
with the constant supersonic velocity **u** since $t = -\infty$. Set the observation
point $P$ on the Mach cone issuing from a source point $S$ located at the
origin of the comoving frame $(x, y, \hat{z})$, and suppose that the contribution of
$S$ towards the sound heard by the observer at the observation time $t_P$ was
made at the retarded time $t = 0$. Any wavelet that is received on the Mach
cone is emitted at the retarded time when the component of the velocity
of the emitting source point in the direction of the observer equalled the
wave speed $c$. It can be seen from the two right-angled triangles in figure 6,
therefore, that the observation time $t_P$ and the coordinate $\hat{z}_P$ of $P$, are in
this case related to the distance $r_P$ of the observer from the trajectory of the
source according to

$$\hat{z}_P = -r_P \cot\theta = -(M^2 - 1)^{\frac{1}{2}} r_P, \qquad (43')$$

and
$$ct_P = r_P \sec\theta = (1 - M^{-2})^{-\frac{1}{2}} r_P, \qquad (44')$$

i.e. according to the relations earlier encountered in (43) and (44).

Figure 6 here

There are other source points, besides $S$, that make their contribution towards the observed field at the retarded time $t = 0$. These lie on a sphere of radius $ct_P$ centered on the observation point $P$ that is in figure 6 designated as $CS$. The wave front emanating—at $t = 0$—from any point on the intersection of this sphere with the source distribution arrives at the observation point $P$ at the observation time $t_P$. The source points which make their contributions towards the observed field at a later instant of the retarded time, $t$, are those that lie on a sphere of radius $c(t_P - t)$ centred on the observation point $P$. In order to find the interval of retarded time during which the extended source in question contributes towards the field at $P$, we must therefore consider a sphere centred at $P$ whose radius decreases with the wave speed $c$ and reduces to zero at $t = t_P$. The observer at $P$ receives contributions only from the interval of retarded time during which such a collapsing sphere intersects the source distribution.

As shown in figure 7, the collapsing sphere $CS$ enters the source distribution considered here at a time $t < 0$, and having passed through the central point $S$ at $t = 0$, it completes sweeping the source at a time $t > 0$. Because the source points in the vicinity of $S$ have a velocity whose component in the direction of the observer is close to $c$, the time taken by the collapsing sphere to cross the source is much longer than the wave-crossing time $L/c$. ($L$ is the length scale of the source distribution along the line of sight of the observer.) It follows from the magnitudes of the angles shown in figure 7 that the difference between the speed of the collapsing sphere and the speed of the source along the radiation direction is given by

$$\Delta u = c - u\cos(\tfrac{1}{2}\pi - \theta + \epsilon) \simeq (1 - M^{-2})^{\frac{1}{2}} \epsilon u, \quad \epsilon \ll 1, \qquad (71)$$

where we have made use of $\theta = \arcsin M^{-1}$ and the fact that $\epsilon$ is much smaller than unity for an observer in the far zone. Therefore, the time taken by the collapsing sphere to cross half of the source is of the order of

$$\Delta t = L/(2\Delta u) \simeq (1 - M^{-2})^{-\frac{1}{2}} L/(2\epsilon u). \qquad (72)$$

To estimate $\epsilon$, let us note that the distance travelled by the source during $\Delta t$ may be calculated in two different ways: it is $u\Delta t$ if inferred from the hypotenuse of the triangle $ABC$ in figure 7, and it is $R_P \epsilon \sec\theta$ if inferred from the length, $R_P \epsilon$, of the side $BC$ of this triangle. [Here $R_P \equiv (r_P{}^2 + z_P{}^2)^{\frac{1}{2}} = ct_P$.] Equating these two values of the traversed distance, and using $\theta = \arcsin M^{-1}$ and (72), we obtain

$$\epsilon \simeq 2^{-\frac{1}{2}}(L/R_P)^{\frac{1}{2}}. \tag{73}$$

This value of $\epsilon$, together with (72), now yields

$$\Delta t \simeq 2^{-\frac{1}{2}}(M^2 - 1)^{-\frac{1}{2}}(LR_P)^{\frac{1}{2}}/c. \tag{74}$$

The reception time interval for the generated signal is the time $L/c$ during which the Mach cones of the various source elements propagate past a stationary observer. Equation (74) states, therefore, that the emission time interval for the signal observed at $P$ exceeds its reception time interval, $L/c$, by a factor of the order of $(R_P/L)^{\frac{1}{2}}$. This factor is huge when the observation point $P$ lies in the far zone.

Figure 7 here

The amplitude of the observed signal can, in turn, be estimated by considering the order of magnitude of the integral that appears in the retarded solution, (3), of the wave equation. The Dirac delta function in (3) expresses the fact that, at any given retarded time, the contributions towards the value of the integral only come from the source points on the collapsing sphere $CS$. So, replacing the spherical-spreading factor $|\mathbf{x} - \mathbf{x}_P|^{-1}$ by the leading term, $R_P^{-1}$, in its expansion in powers of $|\mathbf{x}|/|\mathbf{x}_P|$, and denoting the average value of the source density $s$ by $\bar{s}$, we can see that the right-hand side of (3) scales as

$$\psi \sim \bar{s}\mathcal{A}c\Delta t/R_P, \tag{75}$$

where $\mathcal{A}$ stands for the area of the disc along which the collapsing sphere $CS$ intersects the source distribution and $\Delta t$ is the interval of retarded time during which this sphere travels across the source. Thus, a direct consequence of the elongation of the emission time interval (74) is that the expression

$$\psi \sim 2^{-\frac{1}{2}}(M^2 - 1)^{-\frac{1}{2}}\bar{s}\mathcal{A}(L/R_P)^{\frac{1}{2}} \tag{76}$$

that is implied by (74) and (75) for the amplitude of the emitted wave is correspondingly enhanced by the factor $(R_P/L)^{\frac{1}{2}}$ and so describes a cylindrical decay.

The source in question need not be a source of sound for this to be true; the physical argument given above remains in force also when the source shown in figures 6 and 7 consists of the moving pattern of an electric charge-current distribution and the emission comprises electromagnetic waves. The only step in the above argument which is altered by the requirements of special relativity is that involving the difference between the speed of the collapsing sphere and the speed of the source along the line of sight of the observer. When the source emits electromagnetic waves, we must first estimate the time interval during which the collapsing sphere intersects the source distribution from the point of view of an inertial observer who approaches $P$ along the radiation direction with the speed $v = u\cos(\frac{1}{2}\pi - \theta + \epsilon)$ at the instant at which the collapsing sphere leaves the source (see figure 7). The corresponding time interval in the laboratory frame can then be inferred from this by means of a Lorentz transformation.

Because the speed, $c$, with which the sphere $CS$ collapses is the same in all inertial frames (in the case of light), the collapsing sphere traverses half of the extent, $L_0$, of the source distribution in the moving frame in question during the time interval

$$\Delta t_0 = L_0/(2c). \tag{77}$$

On the other hand, the Lorentz transformation from the moving frame to the laboratory frame implies that†

$$\Delta t = (1 - v^2/c^2)^{-\frac{1}{2}} \Delta t_0 \quad \text{and} \quad L = (1 - v^2/c^2)^{\frac{1}{2}} L_0, \tag{78}$$

in which

$$v \equiv u\cos(\tfrac{1}{2}\pi - \theta + \epsilon) \simeq c - \epsilon u \cos\theta, \quad \epsilon \ll 1. \tag{79}$$

Inserting (77) in (78), we therefore find that $\Delta t$ is of the order of $L/(\epsilon u)$ as in (72). The remainder of the foregoing argument is equally applicable to electromagnetic waves as to sound waves.

So far, we have built up the field of the extended source that we have been considering from the superposition of the fields of the moving source points that constitute it. Alternatively, one can build up the field of an extended moving source from the superposition of the fields of a fictitious set of stationary point sources. The distribution of the required stationary point sources is given by the alternative formulation (25) of the retarded solution to the wave equation. As can be seen from the analogy between (25) and the corresponding solution of the Laplace's equation, the field that is detected by the observer at $\mathbf{x}_P$ at a particular time $t_P$ is the same as that which is produced by a time-independent source distribution with the density $s(\mathbf{x}, t_P - |\mathbf{x} - \mathbf{x}_P|/c)$.

---

† From the time the collapsing sphere enters the source distribution until it reaches the source point $A$ shown in figure 7, the component of the speed of the source along the radiation direction is greater than $c$. For this part of the motion, one must use the superluminal version of the Lorentz transformation (see Barut & Chandola 1993 and the references therein) rather than its usual subluminal version. The relationship between $\Delta t$ and $L$, however, turns out to be the same for both parts of the motion.

The extent of the equivalent stationary source—which changes from one observer to another—can be, in the supersonic (or the superluminal) case, much greater than the extent, $L$, of the actual moving source. This may be seen by considering the distance between the stationary counterparts of any pair of source points with the following positions relative to the observer: a source point $S$ whose caustic passes through the observation point, and a second source point $S'$ with the same value of the longitudinal coordinate $\hat{z}$ ($\equiv z - ut$) which lies within the domain of dependence of the observation point at a distance $l_x$ from $S$.

Figure 8 here

The space-time trajectories of these two superluminally moving source points are two parallel lines and form an angle with the time-axis that, as shown in figure 8, exceeds $\frac{1}{4}\pi$. The space-time locus of the collapsing sphere $CS$ is the past light cone of the observer, a cone whose opening angle with respect to its axis of symmetry equals $\frac{1}{4}\pi$, and whose vertex lies at the observation point $P$. The trajectory of $S$ intersects the past light cone of the observer at two distinct points corresponding to the two retarded times and positions at which the source point $S'$ makes its contribution towards the field at $P$. On the other hand, the two retarded times associated with the source point $S$—whose caustic passes through the observation point—coincide and so the trajectory of $S$ is tangential to the past light cone issuing from $P$. In the lower half of figure 8, the trajectories of $S$ and $S'$ are projected onto a hyperplane $t = $ constant which passes through the point of tangency of the trajectory of $S$ with the past light cone of the observer, i.e. they are projected onto space at the instant at which $S$ makes its contribution towards the field at $P$.

In space-time, the equivalent stationary source distribution is represented by the intersection of the locus (the world tube) of the actual moving source with the past light cone of the observer [see(25)]. The distance between the counterparts, or images, of the source points $S$ and $S'$ within the equivalent stationary source distribution is therefore the distance between points $A$ and $B$ (or $A$ and $C$) of figure 8.† We have already denoted the distance between the source points $S$ and $S'$, i.e. the length $AA'$ of figure 8, by $l_x$. If we also denote the angle subtended by, and the length of, the chord $BC$ by $2\alpha$ and $2l_z$, respectively, then from the triangle $PA'C$ we have

$$l_x = R_P(1 - \cos\alpha) \simeq \tfrac{1}{2}R_P\alpha^2, \quad \alpha \ll 1, \tag{80}$$

and

$$l_z = R_P \sin\alpha \simeq R_P\alpha, \quad \alpha \ll 1, \tag{81}$$

so that

$$l_z \simeq (2l_x R_P)^{\frac{1}{2}}, \qquad l_x/R_P \ll 1. \tag{82}$$

---

† The moving source point $S'$ maps into two stationary source points, but the image of $S$ is unique.

This shows that the difference between the $z$-coordinates of the stationary counterparts of $S$ and $S'$ is by the factor $(2R_P/l_x)^{\frac{1}{2}}$ greater than the actual distance between these two moving source points.

Since $S$ and $S'$ each represent a set of source points, it follows therefore that the $z$-extent of the fictitious time-independent source distribution, whose field mimics the field observed at $(\mathbf{x}_P, t_P)$, is by a factor of the order of $(R_P/L)^{\frac{1}{2}}$ greater than the scale $L$ of the actual moving source distribution. The reason for such a large discrepancy between the scales of the two distributions is the following.

Those source points within the moving source distribution that approach the observer with the wave speed at the retarded time (like $S$), generate a much stronger field at the position of the observer than those whose velocities in the radiation direction are different from c (like $S'$). The only way in which the mathematical formalism can take account of this difference between the radiation efficiencies of $S$ and $S'$ when we build up the field of a (uniformly distributed) moving source from that of a set of stationary source points is to mimic the effect of $S'$, which emits less efficiently, by a stationary source point that lies further away from the observer. Since a measure of the relative radiation efficiencies of the moving source points $S$ and $S'$ is the ratio of the emission to reception time intervals of their signals, this requires that the distances of the stationary counterparts of $S$ and $S'$ relative to the observer should differ by the same factor $(R_P/L)^{\frac{1}{2}}$ that was earlier encountered in (74).

From the point of view of the formulation (25) of the retarded solution to the wave equation, the emission considered here decays cylindrically because the $z$-extent of the stationary counterpart of the source, and hence the range of integration with respect to $z$ in (25), is in the present case of the order of $(R_P L)^{\frac{1}{2}}$. The radiation is beamed also from this point of view because, unless the observer lies on the caustic of one of the moving source points, the stationary counterparts of the various volume elements of the source distribution will not be as widely separated as found here.

Let us next consider a source of the same type which has the finite duration $0 < t < T$, i.e. has the density distribution given in (30). In this case, the collapsing sphere $CS$ only crosses part of the source before the source dies out (see figure 9). As a result, only a limited set of source points—the set that is swept by the sphere $CS$ during $0 < t < T$—contributes towards the signal which reaches $P$ at $t_P$. The extent of the contributing part of the source along the line of sight of the observer, $\Delta L$, can be estimated by setting the interval of retarded time, $\Delta t$, from which contributions are received at $P$ equal to the lifetime of the source, $T$. From expression (74) with $\Delta t \to T$ and $L \to \Delta L$, it then follows that

$$\Delta L \simeq 2(M^2 - 1)(cT)^2/R_P. \qquad (83)$$

The time taken by the set of caustics issuing from this contributing part of the source to propagate past the observer is $\Delta t_P = \Delta L/c$, so that the resulting signal appears as a subpulse of width

$$\Delta t_P \simeq 2(M^2 - 1)cT^2/R_P. \qquad (84)$$

Though composed of waves that are emitted over the entire lifetime of the source, $T$, this subpulse has a width that is by a factor of the order of $cT/R_P$ smaller than $T$.

Figure 9 here

The finiteness of the duration of the source cannot alter the fact that the amplitude of the generated signal decays cylindrically; it only influences the width of the signal. Each element within the contributing part of the source (the shaded region next to the collapsing sphere $CS$ in figure 9, or more precisely Region $I$ of figure 3) emits a set of waves during the lifetime of the source whose fronts later appear as shown in figure 5. In the region next to the envelope of these wave fronts that is covered by two sets of wavelets, i.e. in Region $I_P$ of figure 5, the field is indistinguishable from that of a source point which has an unlimited duration. Provided that the superposition of the Regions $I_P$ of the wave systems of the various contributing source elements takes place in the same way as it does in the case of an infinitely long-lived source, therefore, it would follow that the amplitude of the subpulse that is generated by a finite-duration source should also decay cylindrically.

In the case of an infinite-duration source, the source elements whose contributions superpose coherently at the observation point $P$ are those which lie on the bifurcation surface issuing from $P$ (see figure 1). The mechanism responsible for this is the same as that which renders the electromagnetic radiation from bunches of charged particles coherent: $N$ particles, each with a charge $q$, which approach the observer—along his line of sight— with identical speeds at the retarded time give rise to a radiation intensity proportional to $(Nq)^2$, whereas the intensity of the radiation that arises from the same particles when they have unrelated velocities is only proportional to $Nq^2$. In the present case, the area of the disc across which the bifurcation surface intersects the source distribution (see figure 1) plays the role of $N$. The source elements adjacent to this disc take part in a coherent emission process because they collectively approach the observer along the radiation direction with the wave speed at the retarded time.

The fact that their common speed is the wave speed is not directly relevant to the coherence mechanism which derives from the bunching of these source elements in phase space. Their motion at the wave speed introduces an additional coherence mechanism, one that gives rise to the Doppler contraction of the reception time interval. Not only do the waves that are emitted by the source elements on the bifurcation surface at a given retarded time interfere constructively at the position of the observer, but in addition their fronts crowd together to such an extent that the contributions from an interval $(R_P L)^{\frac{1}{2}}/c$ of retarded time are concentrated into the interval $L/c$ of reception time.

Both of these coherence mechanisms are at work also when the duration of the source is finite. We have already seen how the emission from the entire lifetime of the source, $T$, is Doppler contracted into a reception time interval of the order of $cT^2/R_P$. The other mechanism of coherence, too, is equally effective in the finite as in the infinite-duration case: the area of the disc across which the collapsing sphere associated with the observation point (43)–(44) intersects the source distribution, i.e. the area covered by the set of source points whose contributions superpose coherently, is of the order of $L^2$ in both cases, where $L$ is the length scale of the source distribution. The dimension of the contributing part of the source in the radiation direction, i.e. the thickness of the shaded region next to the collapsing sphere in figure 9, has the small value $(cT)^2/R_P$ but this dimension of the source only influences the duration of the subpulse. The amplitude of the generated signal, which is dictated by the area of the part of the collapsing sphere that lies within the source distribution, is of the same order of magnitude and decays in the same way as that of a signal that is generated by an infinite-duration source—provided, of course, that a dense set of source elements lie adjacent to the collapsing sphere. (For there to be a dense set of source points whose contributions interfere constructively at the observation point the source should, in general, be distributed over a volume.)

The subpulse arising from a finite-duration source is narrower the further away it is observed from the source [see (84)]. In contrast with a corresponding spherically spreading pulse whose amplitude decreases like $R_P{}^{-1}$ and whose temporal width (or spatial width along the direction of propagation) remains constant, the subpulse in question has both a diminishing amplitude ($\sim R_P^{-\frac{1}{2}}$) and a diminishing width ($\sim R_P^{-1}$). The slower decay of its amplitude, combined with the narrowing of its width, still make the energy ($\sim$ amplitude$^2 \times$ width) obey the inverse square law despite the violation of this law by the field strength.

The extended source shown in figure 9, in addition, generates signals which, at the time $t_P$, are detectable at points neighbouring $P$. The collapsing sphere $CS'$ in this figure, for instance, sweeps a different part of of the source during the time interval $0 < t < T$. The propagating caustic arising from the limited part of the source that is swept by $CS'$ also appears to an observer in the far zone as a subpulse of duration $\Delta t_P$, but a subpulse which at the time $t_P$ is detectable at a point different from $P$. The collection of subpulses thus generated by different parts of the source distribution is at any given time present within a pulse window whose width $L/c$ is determined by the length scale $L$ of the source distribution in the radiation direction. The subpulses within this overall waveform can each have an independent identity: each individual subpulse is formed by the accumulation—in an exceedingly narrow region of space—of a set of wave fronts that are emitted by a distinct aggregate of source points over the entire lifetime of the source. Although individually formed by waves which are in phase at the position of the observer, the subpulses that are detected at different locations within the pulse window need not bear any phase relationship to one another. They constitute a (continuous) collection of coherent structures that are superposed incoherently (see figure 10).

Figure 10 here

Because the subpulses become narrower the further away from the source they are observed, the low-frequency content of the spectrum of the radiation field $\psi$ is reduced as $R_P$ increases. The Fourier transform of $\psi$ is the product of the Fourier transforms of the source density $s$ and the Green's function $G_1$, for the right-hand side of (31) has the form of a convolution integral. It is the Fourier transform of $G_1$ that is $R_P$-dependent: the scale length $\Delta l$ of the contributing part of the source, whose diminution with $R_P$ is responsible for the narrowing of the subpulses, is determined by the extent of the portion of the source that is swept by the collapsing sphere. The existence and properties of the collapsing sphere itself is, in turn, dictated by the structure of the Green's function for the free-space wave equation and hence that of $G_1$ [see (32)]. To generate radiation of a given frequency, it is of course necessary that the Fourier transform of the source density should be non-zero at that frequency. However, the singularity associated with the Doppler factor in the expression (32) for $G_1$ renders the high-frequency content of this function so large that the amplitude of the radiation can in the present case be appreciable even at frequencies at which the Fourier components of the source density are small and the radiation that would arise from a conventional emission mechanism is negligible.

The spectrum of a subpulse of width $\Delta t_P$ contains a range of wavelengths whose lower limit is of the order of

$$\lambda \simeq c\Delta t_P \simeq 2(M^2 - 1)(cT)^2/R_P \qquad (85)$$

[see (84)]. If we regard the dimension $uT$ of the region of space swept by the source during its lifetime as that of an effective aperture or antenna which launches the signal, then the Rayleigh distance associated with each subpulse is $L_R = (uT)^2/\lambda$ and so, according to (85), has the value

$$L_R \simeq \tfrac{1}{2}(1 - M^{-2})^{-1} R_P. \qquad (86)$$

The rate of change of the widths of the subpulses with $R_P$ is therefore such that the Rayleigh distance associated with these signals is always of the same order of magnitude as their distance from their source. This means that the subpulses remain at the boundary between the near and far zones permanently and are, as a result, never subject to the spherical spreading that occurs in the far field.

## 5. Discussion

Examples of the type of source that we have analyzed in Sections 3 and 4 may be contained in the turbulent flow of a supersonic jet. Two salient features of the experimental data on such flows, i.e. the level of intensity and the crackling quality of the acoustic radiation that they generate (Ffowcs Williams *et al.* 1975), are consistent with the fact that their distributions satisfy the constraint expressed in (30). The crackle is a possible manifestation of the subpulse structure of the generated waveform: because the subpulses arise from differing parts of the source distribution and are individually coherent, the overall waveform that results from their incoherent superposition can have a highly uneven and rugged amplitude (see figure 10). That the level of intensity of the high-frequency noise can only be explained by the conventional theory if nonlinear propagation effects feed energy into the higher frequencies (Crighton 1986) might reflect the fact that the amplitude of the generated noise decays cylindrically rather than spherically.

The flow in the medium surrounding a supersonic helicopter rotor or propeller is another type of source whose acoustic field shows features consistent with the present analysis. The assumption central to an important part of rotor acoustics that much of the disturbed flow is steady in the blade-fixed frame, i.e. that the variables which enter the expression for the source density $s$ (in the acoustic analogy) are functions of the azimuthal angle $\varphi$ and time $t$ in only the combination $\varphi - \omega t$, implies a rigidly rotating source distribution pattern. Even though the fluid motion around the rotor, that creates the source with the density $s$, has velocities which are usually much smaller than that of sound ($c$), the phase velocity of the pattern associated with the source distribution exceeds $c$ beyond the sonic cylinder $r = c/\omega$. It is in fact the propagation, around the rotation axis, of this pattern of source distribution which imposes the symmetry $\partial/\partial t = -\omega \partial/\partial \varphi$ also on that element of the sound field and so determines the relevant Green's function for the problem: a field with such a symmetry can be built up from that of a point source which moves in a circle with a constant angular frequency $\omega$ (see Ardavan 1989).

On the other hand, the circular trajectories of the constituent volume elements of the distribution pattern in question can be approximated by a set of short rectilinear segments. Since the cylindrically decaying contributions towards the field can only come from those volume elements of the source pattern which move towards the observer with the wave speed and zero acceleration at the retarded time, only a limited set of these short-duration rectilinear trajectories can contribute towards the field at any given observation time. The field generated by a circularly moving source element during the (infinitesimally) short time interval in which it moves along a rectilinear trajectory towards the observer, on the other hand, is not distinguishable from that which is generated by an element of a rectilinearly moving source that is short-lived. One expects, therefore, that the results of the analysis presented in Sections 3 and 4, which are valid for an arbitrarily small value of the source duration, $T$, should be applicable also to a rotating supersonic source, and that the emission from a helicopter rotor or propeller should, in the supersonic regime, have a subpulse structure and a rate of decay that are similar to the Mach wave elements of jet noise. A direct analysis of the rotating problem, though mathematically much more involved than the analysis presented here, fully confirms this expectation (see Ardavan 1994 a).

The counterpart of (1) in the case relevant to the blade-steady sources in a hovering helicopter rotor or propeller is the source density

$$s(r, \varphi, z, t) = s(r, \hat{\varphi}, z) \qquad (87)$$

in which
$$\hat{\varphi} \equiv \varphi - \omega t, \tag{88}$$
and $(r, \varphi, z)$ are the cylindrical polar coordinates based on the axis of rotation. The wave equation (4) under a corresponding symmetry, i.e. under the assumption that the field also depends on $\varphi$ and $t$ in only the combination $\hat{\varphi}$, reduces to

$$\frac{1}{r}\frac{\partial}{\partial r}\left(r\frac{\partial \psi}{\partial r}\right) + \frac{\partial^2 \psi}{\partial z^2} + \left(\frac{1}{r^2} - \frac{\omega^2}{c^2}\right)\frac{\partial^2 \psi}{\partial \hat{\varphi}^2} = -4\pi s(r, \hat{\varphi}, z), \tag{89}$$

which is an equation of the mixed type: it is elliptic in $r < c/\omega$ and hyperbolic in $r > c/\omega$. We can regard the variable $\hat{\varphi}/\omega$ in the domain $r > c/\omega$ as a time coordinate and interpret (89) as the wave equation governing the generation and propagation of (two-dimensional) axisymmetric waves in a non-homogeneous medium for which the wave speed varies like $[1 - c^2/(r\omega)^2]^{-\frac{1}{2}}c$ with the distance $r$ from the axis of symmetry. From this point of view, the fact that the waves propagating along a caustic of the ray conoid of (89) should decay cylindrically becomes a mathematical consequence of the reduction in the dimension of the wave equation that is effected by the symmetry $\partial/\partial t = -\omega \partial/\partial \varphi$ (see Courant & Hilbert 1962).

Figure 11 here

Just as the spherical wavelets emanating from a rectilinearly moving supersonic point source form a Mach cone, so the envelope of the corresponding wavelets from a circularly moving supersonic point source constitutes a caustic (see figure 11). This caustic begins issuing from the point source in the form of a cone with the same opening angle as that of a Mach cone and, after joining a second sheet, eventually develops into a tube-like surface which spirals around the rotation axis to infinity. The two sheets of the caustic are tangent to one another and so form a cusp along the curve where they meet (see the figures in Ardavan 1994 a).

We have seen that the source points which give rise to the cylindrically decaying field in the rectilinear case are those that lie on the bifurcation surface, an inverted cone issuing from the observation point that is the mirror image of the caustic (figure 1). What plays the role of this inverted cone in the present case is a surface in the space of source points which has the same shape and points in the same direction as the reflection of the caustic shown in figure 11 across the meridional plane passing through its conical apex. The source points lying inside this inverted caustic (which we shall again refer to as the bifurcation surface) influence the field at the observation point via waves that, though received simultaneously, were emitted at at least three different values of the retarded time. But the source points outside this surface influence the field at its apex at only a single instant of earlier time. The source points *on* the intersection of the bifurcation surface with the source distribution are the ones which approach the observer with the speed of sound and so generate the constructively interfering waves with tangent rays. Amongst these, the points located on the the cusp curve, where the two sheets of the bifurcation surface meet, approach the observer along the radiation direction not only with the wave speed but also with zero acceleration at the retarded time, and so represent the source points for which the focusing of the rays at the observation point is sharpest. It is in fact the contribution of the source elements whose positions at the retarded time coincide with the cusp curve of the bifurcation surface that gives rise to the cylindrically decaying component of the radiation at the observation point.

The observation points within a finite solid angle—depending on the extent of the supersonic portion of the source distribution—on the opposite sides of the plane of rotation have bifurcation surfaces whose cusp curves intersect the source. The cusp curves belonging to the bifurcation surfaces of different observation points intersect the source distribution along different filaments. Because the signals received by a distant observer at two neighbouring instants in time arise from distinct, coherently radiating filamentary parts of the source which might have both different extents and different strengths, the resulting overall waveform in the present case, too, can consist of an incoherent superposition of a (continuous) set of narrow and coherent subpulses. The energy carried by the subpulses falls off as the inverse square of $R_P$ as in the rectilinear case: the widths of the subpulses decrease like $R_P{}^{-1}$ with the distance $R_P$ from the source and this combines with the fact that their amplitudes decay like $R_P{}^{-\frac{1}{2}}$ rather than $R_P{}^{-1}$ to give the $R_P{}^{-2}$ dependence. Each subpulse behaves like an individual missile and the filament responsible for it acts as an independent source, so that the volume distribution of sources may be thought of as the linear superposition of a collection of continually operating missile launchers each of which points in a different direction.

The probable crackling associated with this subpulse structure might be the explanation for an effect that has been experimentally known since the early works of Bryan (1920) and Hilton (1939); crackle abounds in the emission from supersonic rotors. However, we have not been able to locate any experimental data on the decay rate of this type of emission in the far zone. In addition to the conventionally studied component of the propeller noise which decays spherically, there is in some special zones a non-spherically decaying component that dominates it at large range. This is a prediction of the present analysis that can be experimentally tested and may well provide the explanation for puzzling features of the already existing data.

An electromagnetic emission whose features closely resemble those of the emission considered here is the radiation received from pulsars (see Lyne & Graham-Smith 1990). It is well established that the periodicity of the signals emitted by a pulsar stems from the spinning motion, about a fixed axis, of a highly magnetized central neutron star. This, in conjunction with the fact that the average waveforms received from pulsars have periods that are remarkably constant, means that the profiles of the pulses in question, when observed in an inertial frame marked by the cylindrical polar coordinates $(r_P, \varphi_P, z_P)$ whose $z_P$-axis is coincident with the spin axis of the neutron star, vary both temporally $(t_P)$ and azimuthally $(\varphi_P)$ as functions of the single variable $\varphi_P - \omega t_P$ with a practically constant $\omega$. In other words, the polar diagram of the radiation emitted by a pulsar has a pattern that rigidly rotates with the angular velocity $\omega$ of the central neutron star of the pulsar, and the electric and magnetic fields constituting this radiation have cylindrical components that possess the symmetry

$$\frac{\partial}{\partial t_P} + \omega \frac{\partial}{\partial \varphi_P} = 0, \qquad (90a)$$

i.e. that are of the functional form

$$f(r_P, \varphi_P, z_P, t_P) = f(r_P, \hat{\varphi}_P, z_P) \quad \text{and} \quad \hat{\varphi}_P \equiv \varphi_P - \omega t_P, \qquad (90b)$$

and so reduce to zero under the action of the differential operator in (90 a).

A radiation with such a polar diagram can only arise from a source whose pattern of distribution has, likewise, a rigidly rotating motion with an angular velocity $\omega$: if the fields possess the above symmetry, then Maxwell's equations require that the densities of the electric charges and currents that produce these fields must possess the same symmetry, i.e. must depend on the (source) coordinates $\varphi$ and $t$ in only the combination $\hat{\varphi} = \varphi - \omega t$. Thus the overall distribution of the plasma within the emitting region of the magnetosphere is required—by the observational data—to have a rotating pattern whose points everywhere move with the linear phase speed $r\omega$; although, of course, none of the charged particles that constitute the source, and provide the medium for the propagation of the pattern associated with the source, are themselves constrained to be corotating (see e.g. Ardavan 1981, da Costa & Kahn 1985).

The linear speed $r\omega$ of such a pattern exceeds $c$, the speed of light *in vacuo*, in the outer part of the pulsar magnetosphere, for in pulsars the light cylinder $r = c/\omega$ is sufficiently close to the central neutron star to lie within its surrounding plasma. The fact that the speed of such a pattern exceeds the velocity of light beyond the light cylinder does not, however, in any way violate the requirements of special relativity. The superluminally moving pattern is created by the coordinated motion of aggregates of subluminally moving particles. Symmetry (90) only demands that the (subluminal) motion of the charged particles in the (mixed) magnetospheric plasma create a pattern—of charge separation, for instance—whose *phase* velocity exceeds the velocity of light when $r > c/\omega$.

The density of the source distribution in the magnetosphere of a pulsar is therefore of the same form as that given in (87) and the equation governing, e.g. the $z$-component of, the electromagnetic potential in the Lorentz gauge is identical to (89). There is, as a result, a strong analogy between the features of the electromagnetic radiation that is received from pulsars and those of the acoustic radiation that is produced by supersonic helicopter rotors and propellers (cf. Schmitz & Yu 1986 and Lyne & Graham-Smith 1990).

From the Maxwell's equations and the symmetry (87) alone there follows a broad-band coherent radiation that is, in all its salient features, similar to the radiation received from pulsars (Ardavan 1994 b). The emitted radiation is beamed into a finite solid angle that depends on the extent of the source distribution and that points in a direction normal to the axis of rotation. At any given observation point within this solid angle, the electromagnetic field arises almost exclusively from those volume elements of the source whose positions at the retarded time match the positions of the set of rigidly rotating points that approach the observer with the wave speed and zero acceleration in the radiation direction. These are the points—collectively forming the cusp curve of the bifurcation surface—at which the Green's function for the problem is most singular. Because the signals received at two neighbouring instants in time thus arise from distinct filamentary parts of the source, the resulting overall waveform again closely resembles that shown in figure 10: the narrow signals which are responsible for the crackle in the acoustic case here appear as what, in the observational literature on pulsars, are called micropulses. Each micropulse embodies a caustic and hence has an amplitude that does not obey the spherical spreading law: its flux density falls off like $R_P^{-1}$, rather than like $R_P^{-2}$, with the distance $R_P$ from the source. The radiation therefore has a brightness temperature that is, by a factor of the order of $R_P\omega/c$, greater than the kinetic temperature of the plasma that generates it, and so is highly coherent.

The spectrum of this radiation in general extends over a wide range of frequencies from radio waves to gamma-rays: since it entails caustics, at which the wave fronts crowd together to such an extent that the wavelength of the radiation is Doppler-shifted to zero, the Green's function for the present problem has a Fourier transform that, in contrast to that of the Green's function for a subluminally moving source, falls off algebraically rather than exponentially at high frequencies.

Furthermore, this radiation consists of two concurrent elliptically polarized modes whose position angles are approximately orthogonal and individually vary across the waveform in the course of each rotation: the phase shift that is generally present between the waves on opposite sides of a caustic in this case endows the polarization state of each micropulse with two distinct values of the polarization angle, values that directly depend on the average orientation of the electric current density along the filamentary site of emission.

We thank D. Lynden-Bell who has read the manuscript and has made helpful comments, and our several colleagues who have borne with us over the long time that the ideas presented here were evolving.

**Figure captions**

Figure 1. The dotted region designates the volume of the $(x, y, \hat{z})$-space occupied by the source distribution. The spherical wavelets emanating from a given source point $S$ and the caustic constituting their envelope are shown by full lines, and the bifurcation surface associated with an observation point $P$ is shown by broken lines. The hatched region next to, and inside, the bifurcation surface represents the part of the source that gives rise to the cylindrically decaying signal observed at $P$.

Figure 2. Intersection of the caustic issuing from the source point $S$ with a plane $z = $ constant. The cone has an opening angle $\theta = \arcsin(c/u)$ and the propagation speed $dl/dt = c$. The radius $r$ of the circle along which the intersection occurs increases with a speed $c_*$ that is supersonic (or superluminal).

Figure 3. The domain of dependence of the observation point $P$ in the $(x, y, \hat{z})$-space. The dotted area represents the volume occupied by the source, the two circles represent the spheres (33) and (34), and the broken lines represent the cone (12). The region shown in black (Region $I$) lies inside the cone (the bifurcation surface) but outside the two spheres, while the hatched region (Region $II$) consists of the union of the two spheres less their intersection. The points $A, B$ and $C$ represent the circles, normal to the plane of the figure, along which the spheres (33) and (34) are tangent to the cone (12) or intersect one another.

Figure 4. The closed contour used for the evaluation of the integral in (68). The segment $C$ runs along the real axis of the $\zeta$-plane from the saddle point $\zeta = 0$ to the boundary point $\zeta = \epsilon$, and $C_1$ and $C_2$ are the constant-phase contours of steepest descents passing through $\zeta = 0$ and $\zeta = \epsilon$, respectively.

Figure 5. The spherical wave fronts emanating from an infinitesimal volume element $S$ of a finite-duration source during its lifetime $0 < t < T$. The points $S_0$ and $S_T$ designate the retarded positions of the source point $S$ at $t = 0$ and $t = T$. The region next to the truncated cone constituting the envelope of these wave fronts, whose points receive contributions from two different retarded positions of the source simultaneously, is Region $I_P$.

Figure 6. The caustic issuing from the origin, $S$, of the comoving coordinate frame $(x, y, \hat{z})$ at the observation time $t_P$. The source distribution is shown at both the retarded time $t = 0$ (on the left) and the observation time

$t = t_P$ (on the right). The collapsing sphere $CS$, centred on the observation point $P$, is shown at the instant ($t = 0$) at which it passes through the source point $S$.

Figure 7. The same as figure 6 with the addition of the configuration of the system at the instants at which the collapsing sphere $CS$ enters ($t < 0$) and leaves ($t = \Delta t > 0$) the source distribution (schematic). The hypotenuse $AB$ of the right-angled triangle $ABC$ has the length $u\Delta t$, and the side $BC$ of this triangle is approximately equal to $R_P \epsilon$ when $R_P \gg L$ and so $\epsilon \ll 1$.

Figure 8. A space-time diagram only showing the spatial dimensions $x$ and $z$. The upper half of the figure depicts the past light cone of the observer $P$ together with the trajectories of two supersonically (or superluminally) moving source points: one ($S$) whose trajectory is tangent to this light cone at the point $A$, and another ($S'$) whose trajectory crosses the cone at the points $B$ and $C$. A hyperplane $t = $ constant passing through $A$ intersects the past light cone of $P$ along the collapsing sphere ($CS$) shown in the lower half of the figure. The projections of the trajectories of the source points $S$ and $S'$ onto the hyperplane $t = $ constant are also shown.

Figure 9. The modified version of figure 7 for a source of finite duration (schematic). During the time interval $0 < t < T$ in which the source is active, the collapsing sphere $CS$ sweeps only the thin portion of the source distribution that is here designated by the hatched region adjacent to this sphere. The part of the source which is swept by another collapsing sphere, $CS'$, has the same width but a different identity.

Figure 10. An example of the overall waveform: a pulse of width $L/c$ which is formed by the incoherent superposition of coherent subpulses of widths $\Delta t_P$. This is one of the radio pulses received from the pulsar PSR 0950+08 at 430 MHz (Cordes & Hankins 1979); the dotted curve is the average of several hundred single pulses.

Figure 11. The envelope of the spherical field wavelets emanating from a supersonically (or superluminally) moving source point in circular motion. The heavier curves show the cross section of the envelope with the plane of the orbit of the source. The larger of the two broken circles designates the orbit and the smaller the sonic (or light) cylinder $r = c/\omega$.